\documentclass{jfm}

\usepackage{graphicx}
\usepackage{natbib}
\usepackage{psfrag}
\usepackage{bm}% bold math
\usepackage{color}
\usepackage{amsmath}
\usepackage{amsfonts}
\usepackage{natbib}
\usepackage{psfrag}
\usepackage{float}
\usepackage{pstricks}
\usepackage{color}

\ifCUPmtlplainloaded \else
  \checkfont{eurm10}
  \iffontfound
    \IfFileExists{upmath.sty}
      {\typeout{^^JFound AMS Euler Roman fonts on the system,
                   using the 'upmath' package.^^J}%
       \usepackage{upmath}}
      {\typeout{^^JFound AMS Euler Roman fonts on the system, but you
                   dont seem to have the}%
       \typeout{'upmath' package installed. JFM.cls can take advantage
                 of these fonts,^^Jif you use 'upmath' package.^^J}%
      }
  \else
  \fi
\fi

\ifCUPmtlplainloaded \else
  \checkfont{msam10}
  \iffontfound
    \IfFileExists{amssymb.sty}
      {\typeout{^^JFound AMS Symbol fonts on the system, using the
                'amssymb' package.^^J}%
       \usepackage{amssymb}%

      }{}
  \fi
\fi

\ifCUPmtlplainloaded \else
  \IfFileExists{amsbsy.sty}
    {\typeout{^^JFound the 'amsbsy' package on the system, using it.^^J}%
     \usepackage{amsbsy}}
    {\providecommand\boldsymbol[1]{\mbox{\boldmath $##1$}}}
\fi

\newsavebox{\astrutbox}
\sbox{\astrutbox}{\rule[-5pt]{0pt}{20pt}}

\newcommand{\vel}{\mathbf{u}}

\newcommand{\ha}{H\!a}
\newcommand{\opl}{\mathcal{L}}
\newcommand{\velD}{\bar{\mathbf{u}}}
\newcommand{\tH}{\tau_H}
\newcommand{\tJ}{\tau_J}
\newcommand{\tD}{\tau_{2D}}
\newcommand{\tQ}{\tau_{Q2D}}

\title[The decay of wall-bounded MHD turbulence at low $Rm$]{The decay of wall-bounded MHD turbulence at low $Rm$}

\author[K. Kornet and A. Poth\'erat]%
{K\ls A\ls C\ls P\ls E\ls R\ns K\ls O\ls R\ls N\ls E\ls T\ns A\ls N\ls D\ns A\ls L\ls B\ls A\ls N\ns  P\ls O\ls T\ls H\ls \'E\ls R\ls A\ls T%
  \thanks{Email address for correspondence: alban.potherat@coventry.ac.uk} 
}

\affiliation{Applied Mathematics Research Centre, Coventry University,
Coventry, CV51FB, UK\\[\affilskip]
}
\pubyear{}
\volume{}
\pagerange{}
\date{?; revised ?; accepted ?. - To be entered by editorial office}
\begin{document}

\maketitle

\begin{abstract}
We present Direct Numerical Simulations of decaying Magnetohydrodynamic (MHD) 
turbulence at low 
magnetic Reynolds number. The domain considered is bounded by periodic boundary
 conditions in the two directions perpendicular to the magnetic field and by 
two plane Hartmann walls in the third direction. High magnetic fields 
(Hartmann number of up to 896) are considered thanks to a numerical method 
based on a spectral code using the eigenvectors of the dissipation operator.
It is found that the decay proceeds through two phases: first, energy and 
integral lengthscales vary rapidly during a two-dimensionalisation phase 
extending over about one Hartmann friction time. During this phase, the 
evolution of the former appears significantly more impeded by the presence of 
walls than that of the latter. Once the large scales are close to quasi-two dimensional, the decay results from the 
competition of a two-dimensional dynamics driven by dissipation in the Hartmann 
boundary layers and the three-dimensional dynamics of smaller scales. 
In the later stages of
the decay, three-dimensionality subsists under the form of barrel-shaped 
structures. A purely quasi-two dimensional decay dominated by friction in the 
Hartmann layers is not reached, because of residual dissipation in the bulk. 
However, this dissipation is not generated by the three-dimensionality 
that subsists, but by residual viscous friction due to 
horizontal velocity gradients. Also, the energy in the velocity component 
aligned with the magnetic field is found to be strongly suppressed, as is 
transport in this direction. This results reproduces the experimental findings 
of \cite{kolesnikov1974_fd}.

\end{abstract}


\section{Introduction}
This work concerns the decay of MagnetoHydroDynamic turbulence in electrically conducting fluids subjected to an externally imposed magnetic field. We are particularly interested in the influence of solid, electrically insulating walls on this process.  This generic problem is relevant to a number of practical engineering problems in the metallurgy and nuclear industry, but also bears relevance to 
some aspects of the dynamics of liquid planetary cores and the associated dynamo problem.
%, in conditions where the flow isn't electrically conductive nor intense enough to induce magnetic fields comparable to an externally imposed one. Such flows are described within the low Magnetic Number approximation (see \cite{roberts67}). 
%We shall further assume that magnetic dissipation acts too fast for Alfv\`en waves to exist (this is the limit of Low Lundquist numbers).  

%how MHD turbulence decays 
When the magnetic field $B\mathbf e_z$ is imposed (in the sense of the low magnetic Reynolds number 
approximation (\cite{roberts67}), turbulence evolves as a result of the 
competition between inertia and the diffusion of momentum along the direction 
of the magnetic field.
A structure of size $l_\perp$ becomes elongated over a length $l_z$ by this 
diffusion over a timescale of $\tau_J (l_z/l_\perp)^2$ (\cite{sm82}), whilst 
loosing 
energy through Joule dissipation ($\tau_J=\rho/(\sigma B^2)$ is the Joule dissipation time, $\rho$ and $\sigma$ are the fluid density and electric conductivity.).  
\cite{moffatt67} first showed that under this linear phenomenology, the 
turbulent kinetic energy decayed at $E\sim t^{-1/2}$ towards an asymptotic 
state where the flow quantities did not vary along the magnetic field (in this 
sense,a \emph{two-dimensional} state) but where the kinetic energy of the 
component along the magnetic field was a third of the total kinetic energy (for 
a three-component flow). This phenomenology was recovered by \cite{schumann76}: 
this author conducted low-resolution direct numerical simulations to confirm 
that this linear phenomenology applied during less than $\tau_J$ but that 
non-linear effects subsequently led to a decay of the kinetic energy associated to the velocity along 
$\mathbf B$. However, he also found 
that the skewness tended to a finite value in the later stages of the decay, 
which he attributed to the persistence of transport of kinetic energy 
$\mathbf B$.
%in the interpreted as the signature of a strong kinetic energy corresponding to the third velocity component. 
%By contrast, \cite{kolesnikov1974_fd} showed that strong scalar transport 
%subsisted across the magnetic field but was suppressed across it, 
%in an experiment where turbulence decayed behind a rectangular grid placed in a rectangular channel.\\
The more recent and higher resolution simulations of \cite{burattini10_jfm}, 
 confirmed Schumann's findings.
%that Moffatt's phenomenology held only during less than the turnover time of the large scales.
%, who also pointed out that the linear theory held better for some physical quantities than others.
Both studies analysed cases where interaction parameter $N=\tau_U/\tau_J$ spanned a range between 0.1 and 50 ($\tau_U(l)=l/U(l_0)$ is the eddy turnover time based on the initial size and velocity of the large scales $l_0$ and $U(l_0)$). 
Using the invariance of the "parallel" component of Loitsyansky's integral $I_\|$, \cite{okamoto10_jfm} showed that for $N>>1$, the $t^{-1/2}$ law for the decay of kinetic energy was recovered and that the integral lengthscale in the direction of the magnetic field evolved as $l_z\sim t^{1/2}$. At moderate values of $N$, a similar approach led the authors to conclude that energy decayed  as $E\sim t^{-11/7}$ while the integral lengthscales along and across the magnetic field increased respectively as $l_z\sim t^{5/7}$ and $l_\perp\sim t^{3/14}$.
These theoretical scalings as well as the invariance of $I_\|$ were verified by 
means of direct numerical simulations at the highest resolution available to date (up to $2048^3$), for $N<1$.\\
Aside of inertia, a second major factor is likely to interfere with Moffatt's 
linear theory: the presence of walls, and in particular Hartmann walls, that 
are perpendicular to the magnetic field. These are indeed a feature of 
practically any of the situations where low $Rm$ MHD turbulence is likely to be 
found. A strictly two-dimensional state is not possible in their presence 
because of the very thin Hartmann boundary layers that develop along them (see 
for instance \cite{moreau90}). Instead, \cite{sm82} theorised that in a channel of width $L$,  a structure of size $l_\perp$ became quasi-two-dimensional after 
$\tau_{2D}(l_\perp)\sim \tau_J (L/l_\perp)^2$. Past this stage, electric current 
in the core became of order $\ha^{-1}$, the ratio of the boundary layer 
thickness to $L$: dissipation occurs then almost exclusively in the boundary 
layers and is equally viscous and magnetic. In contrast, strictly 
two-dimensional states are possible when walls are absent and the Joule 
dissipation can therefore drop to much lower values. \cite{kolesnikov1974_fd} 
also observed experimentally that in the presence of Hartmann walls, transport along the magnetic field was suppressed in the later stages of the decay. This was interpreted as an evidence of suppression of the velocity component in this direction, in contrast to the prediction of theories and simulations where no wall was present. 
Nevertheless, although a "through" velocity component is precluded by the walls, Ekman pumping can still potentially lead to strong vertical 
velocities at moderate $N$ \citep{psm00_jfm}. More recently, 
it was also found that even for $N>1$, a small amount of three-dimensionality 
could lead to a complex system of three-dimensional co- and contrarotating 
recirculations \citep{prcd13_epje,bpd15_jfm}.\\
Until recently, numerically simulating MHD turbulence at high $N$ in the presence of walls 
incurred prohibitive computational costs because of the need to resolve the 
Hartmann boundary layers. Recently, the authors took a different approach to the
simulation of these flows based on spectral methods using bases of functions 
whose elements already incorporate these layers. 
These partly alleviate this computational constraint to the point where the 
computational cost becomes independent of $\ha$ (\cite{dyp09_tcfd,kop15_jcp}). 
We propose to take advantage of this new technique to investigate decaying turbulence in a channel bounded by electrically insulating walls in view of answering the following 
questions.
\begin{enumerate}
\item Does three-dimensionality subsist in the later stages of the decay ($t>>\tau_{2D}(l_\perp)$)?
\item Which part of the energy subsists in the third velocity component ?
\item How do Hartmann walls affect the early phases of the decay ($t<\tau_{2D}(l_\perp)$) ?
\end{enumerate}
We shall first recall the governing equations and the timescales that govern the problem (section \ref{sec:eq_theor}). Our numerical method and simulation strategy is presented in section \ref{sec:num}. We then examine the earlier decay phase which is expected to present the strongest similarities with earlier works not involving walls (section \ref{sec:3ddecay}). The later stages of the decay where similarities with two-dimensional turbulence are expected are 
analysed in section \ref{sec:2ddecay}.  The robustness of our results is tested  in section \ref{sec:bigbox} by changing initial conditions and domain size.
%question  the nature of the 2D state with walls
%Ekman pumping makes 3rd component still possible 
%complex structure
%non-locality of MHD turbulence: can the walls influence the decay of turbulence before vortices reach them ?

\section{Governing equations \label{sec:eq_theor}}
\subsection{Problem definition}
\label{s:goveq}
At low Magnetic Reynolds number, the full system of the induction equation and 
the Navier-Stokes equations for an incompressible fluid can be approximated to 
the first order (The Magnetic Reynolds number $R\!m$ represents the ratio of the induced magnetic field to the imposed one). This leads to the following system \citep{roberts67}:
\begin{eqnarray}
\frac{\partial \vel}{\partial t} + (\vel \cdot \nabla) \vel  &=& 
- \frac{1}{\rho} \nabla p + \nu \Delta\vel  + \frac{1}{\rho} \mathbf{j} \times \mathbf{B} \,, \label{eq:nav}\\
\nabla \cdot \vel &=& 0 \,, \label{eq:cont}\\
\nabla \cdot \mathbf{j} &=& 0 \,, \label{eq:contj}\\
\mathbf{j} &=& \sigma ( - \nabla \Phi + \vel \times \mathbf{B}) \,, \label{eq:ohm}
\end{eqnarray}
where $\vel$ denotes fluid velocity, $\mathbf{B}$ - externally imposed magnetic
field, $\mathbf{j}$ - electric current density, $\nu$ - kinematic viscosity,
$\sigma$ - electrical conductivity, $\Phi$ - electric potential.  We consider a
channel flow with a homogeneous transverse magnetic field $B\mathbf e_z$ and
impermeable ($\vel|_{wall} = \mathbf{0}$), electrically insulating ($\mathbf{j
\cdot n}|_{wall} = \mathbf{0}$) walls located at $z=\pm L/2$ (see fig.~\ref{fig:channel}).  In the $xy$
directions we impose periodic boundary conditions with period $L$. Following 
\cite{roberts67}, the Lorentz force can be expressed as the sum of a
gradient of magnetic pressure $p_m$ and a rotational term:
\begin{figure}
\centering
\psfrag{U}{$\mathbf U$}
\psfrag{B}{$\mathbf B$}
\psfrag{ex}{$\mathbf e_x$}
\psfrag{ey}{$\mathbf e_y$}
\psfrag{ez}{$\mathbf e_z$}
\psfrag{al}{$\alpha$}
%\psfrag{2L}{$L$}
\psfrag{L}{$L$}
\includegraphics[width=8cm]{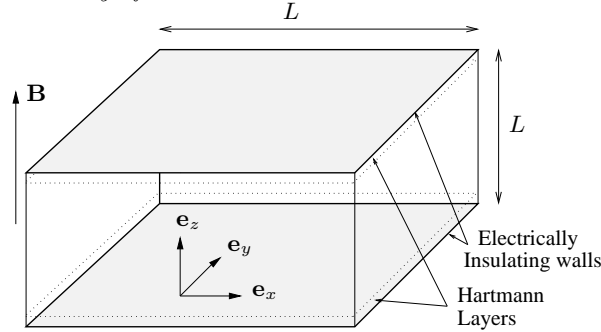}
\caption{Geometry of the Channel flow with transverse magnetic field}
\label{fig:channel}
\end{figure}
\begin{equation}
\mathbf{j} \times \mathbf{B} = -\nabla p_m  - \sigma B^2 \Delta^{-1}\partial_{zz} \vel.
\end{equation}
Using the above identity and adopting  the reference scale $L$, time $L^2/\nu$
and velocity $\nu/L$ the set of equations (\ref{eq:nav}-\ref{eq:ohm}) can be
expressed in dimensionless form:
\begin{equation}
\frac{\partial \vel}{\partial t} + P[(\vel \cdot \nabla) \vel] = 
\Delta\vel - \frac{1}{\ha^2} \Delta^{-1}\partial_{zz} \vel \,,
\label{invlapl}
\end{equation}
where $\ha=LB\sqrt{\sigma/\rho\nu}$ is the Hartman number and $P$ denotes
orthogonal projection onto the subspace of solenoidal fields. 
\subsection{Timescales of the decay \label{sec:timescales}}
We now consider an initially turbulent, isotropic flow, left to decay from 
$t=0$ under the action of the Lorentz force and viscous friction, in the 
configuration described above. Hartmann walls can be expected to exert little 
influence on the initial phase of the decay, during which strong
two-dimensionalisation of the flow should take place, as in the simulations of 
\cite{schumann76,okamoto10_jfm} and others. Unlike, in this previous studies, the presence of physical walls may lead to a physically realistic phase 
beyond this initial one, where the flow dynamics may become a two-dimensional 
to some extent. To obtain a first estimate for the timescale for the 
transition between these two phases, let us first consider a single 
 turbulent structure of size $l_\perp$, which diffuses across the channel 
width $L$ under the action of the Lorentz force in time 
$\tau_{2D}(l_\perp)=\tJ(L/l_\perp)^2$. If this timescale is shorter than both 
the inertial timescale $\tau_U(l_\perp)=l_\perp/U(l_\perp)$ and the viscous 
timescale $\tau_\nu(l_\perp)=\nu/l_\perp^2$, then this single structure is 
quasi-two-dimensional. Two conditions for two-dimensionality ensue:
\begin{eqnarray}
\left(\frac{l_\perp}{L}\right)^3&>&\frac{l_0}{L}\frac{U(l_\perp)}{U(l_0)}\frac1{ N(l_0)}\label{eq:2d_cond_n},\\
\left(\frac{l_\perp}{L}\right)^2&>&\frac1\ha,
\label{eq:2d_cond_nu}
\end{eqnarray}
where $l_0$ denotes the size of the large scales at $t=0$. Since both these 
conditions are scale-dependent, each of them defines a minimum two-dimensional 
scale. Consequently, in a turbulent flow where 
$N=\tau_U(l_\perp)/\tJ>>1$, larger scales are two-dimensional while smaller 
scales may still be three-dimensional.
From inertial condition (\ref{eq:2d_cond_n}), the smallest quasi-two-dimensional 
structure satisfies $l_\perp^{2D}\sim L [(l_0U(l_\perp^{2D})/(N(l_0)LU(l_0))]^{1/3}$. Since the two-dimensionalisation time $\tD(l_\perp)$ increases with 
$l_\perp$, this scale is also the slowest structure to become two-dimensional 
and so $\tD^M=\tD(l_\perp^{2D})$ provides an estimate for the two-dimensionalisation phase: 
\begin{equation}
\tD^M\sim\tJ \left(N(l_0) \frac{U(l_0)h}{U(l_\perp^{2D})l_0}\right)^{2/3}.
\label{eq:t2dinert}
\end{equation}
$\tD^M$ depends on $U(l_\perp^{2D})$, which is expected to drop by several 
orders of magnitudes during the two-dimensionalisation phase. 
Not only does this 
variation of $U(l_\perp^{2D})$ considerably slow down two-dimensionalisation 
at the initial scale $l_\perp^{2D}(t=0)$ , but scales smaller than the initial value of $l_\perp^{2D}$ may satisfy 
(\ref{eq:2d_cond_n}) %as the value of $U(l_\perp)$ decreases, which 
and may in turn become quasi-two-dimensional 
in a time significantly longer than the initial value of (\ref{eq:t2dinert}). On these grounds, 
a lower estimate for the two-dimensionalisation time is obtained by evaluating 
(\ref{eq:t2dinert}) based on the value of $U(l_\perp)$ at $t=0$ (this value 
is fixed by the choice of initial turbulent spectrum).

From viscous condition (\ref{eq:2d_cond_nu}), by contrast, the smallest quasi-two-dimensional scale $l_\perp^{2D}$ does not depend on $U(l_\perp)$, and neither 
does the associated two-dimensionalisation time:
%$\tau_{2D}(l_\perp^\nu)\sim\nu/l_\perp^{2D}$,  
%
\begin{eqnarray}
l_\perp^{2D}\sim L\ha^{-1/2} \label{eq:l2d_nu}\\
\tD^M=\tau_{2D}(l_\perp^{2D})\sim\tJ\ha^{-1}=\tH \label{eq:t2d_nu}
\end{eqnarray}
Since no scale smaller than (\ref{eq:t2d_nu}) can become two-dimensional 
regardless of how much turbulent intensity drops, $\tH$ represents a closer 
estimate of the timescale for two-dimensionalisation of the whole turbulent 
flow than (\ref{eq:t2dinert}).
\subsection{Two-dimensional decay \label{sec:2d_decay}}
Once all structures have become quasi two-dimensional, the evolution of the 
flow is governed by two-dimensional dynamics with an added friction due to 
the Hartmann layers. \cite{sm82} showed that the velocity averaged 
across the channel $\bar{\mathbf u}=\int_{-1}^1 \mathbf u_\perp dz$ satisfied a shallow water equation of the form:
\begin{eqnarray}
\frac{\partial \velD}{\partial t} + (\velD \cdot \nabla_\perp) \velD  &=&
- \frac{1}{\rho} \nabla_\perp p + \nu \Delta_\perp\velD  - \frac{2 \velD}{\tH}\,,\label{eq:sm82}\\
\nabla_{\perp} \cdot \velD &=& 0, \,
\end{eqnarray}
where operators with subscript $\perp$ operate in the $x$-$y$ plane only. 
$\tH$ appears as the typical time for the dissipation due to the Hartmann boundary layers, and therefore a characteristic time of the two-dimensional dynamics. 
From (\ref{eq:sm82}), the evolution of the total kinetic energy $E\simeq E_{2D}=\|\velD\|^2_{2}$ associated to the mean flow $\velD$ when the flow 
follows a two-dimensional dynamics reduces to
\begin{equation}
\frac12\frac{dE_{2D}}{dt}=-2\frac{E_{2D}}{\tH}-\nu\|\nabla\velD\|^2_{2D},
\end{equation}
where $\|\cdot|_{2D}$ represents the two-dimensional $\mathcal L^2$ norm.
Introducing lengthscale $l_\perp^\nu=(\|\velD\|^2_{2D}/\|\nabla\velD\|^2_{2D})^{1/2}$ which characterises velocity gradients in the $(x,y)$ plane, it comes that
\begin{equation}
\frac{dE_{2D}}{dt}=-\frac4\tH\left[1+\frac1{2\ha}\left(\frac{L}{l_\perp^\nu}\right)^2\right]E_{2D}.
\label{eq:2d_energy}
\end{equation}
%
%The first term is of the order of $\ha (l_\perp^\nu/h)^2$, where $l_\perp^\nu$ 
%is the typical scale of velocity gradients in the $(x,y)$ plane. 
It follows from the respective definitions of the total kinetic energy $E$ and $E_{2D}$, that 
\begin{equation}
E=E_{2D}\left(1+\mathcal O(\max\{\ha^{-1},\alpha^2\})\right), 
\end{equation}
where $\alpha=\|\mathbf u-\velD\|/\|\mathbf u\|$ represents the degree of 
three-dimensionality in the flow. For a quasi-two-dimensional flow, 
$E=E_{2D}(1+\mathcal O(\ha^{-1}))$. We shall see from the analysis of flow profiles in section \ref{sec:2ddecay}  (figure \ref{fig:maxu})
that in the later stages of the decay, $\alpha\lesssim 0.1$ so $E_{2D}$ 
can be expected to provide a good approximation for $E$ in the two-dimensional 
phase of the decay.\\
The two-dimensional dynamics of the flow is expected to favour the formation of 
large scales and indeed \cite{schumann76} showed that energy transfer towards 
them occurred during the decay. Areas of strong shear may however persist between them. 
Furthermore, the typical lengthscale of the viscous core of 
quasi-two-dimensional MHD is known to scale as $L Ha^{-1/2}$ \citep{sommeria88}. For such fine 
quasi-two dimensional structures, Hartmann friction and horizontal viscous 
friction would be of the same order. For large structures on the other hand, 
$l_\perp^\nu/L$ should be of the order of unity and so in the limit 
$\ha\rightarrow\infty$, the decay should be 
strongly dominated by Hartmann friction. The total kinetic energy should then 
decay as $E\sim \exp(-4t/\tH)$. 
Any discrepancy to exponential decay of this form is therefore the signature 
either of thin quasi-two-dimensional structures or of a residual 
three-dimensionality. We shall attempt to measure this 
discrepancy in our numerical simulation to identify the mechanisms of the
 long-term decay. It should, however be noted that for structures such 
that $l_\perp^\nu/L\sim Ha^{-1/2}$, three-dimensionality subsists anyway 
because at this scale,  friction between horizontal planes balances the 
diffusion of momentum along the magnetic field.
\section{Numerical approach\label{sec:num}}
\subsection{Numerical method}
The problem set out in section \ref{s:goveq} is solved numerically, using a new
type of spectral method designed to alleviate the computational cost associated with strong anisotropy and thin Hartmann boundary layers. Thanks
to it, increasing the magnetic field incurs essentially no direct computational 
cost per time step.  The mathematical foundations of this method and
numerical implementations are described in detail in
\cite{dyp09_tcfd,kop15_jcp},  where it is also tested for the exact channel 
geometry studied here. For the sake of completeness, we shall nevertheless 
outline the principle of this new method.  Using the spectral approach we seek
the solution of eq. (\ref{invlapl}) as the decomposition on elements of basis
$\mathbf{u}_i$:
\begin{equation}
\mathbf{u} = \sum_i c_i(t) \mathbf{u}_i(\mathbf{x}). \label{eq:sp}
\end{equation}
As the spatial dependence is carried solely by $ \mathbf{u}_i$, when
representation (\ref{eq:sp}) is injected into eq. (\ref{invlapl}), the latter
reduces to the set of ordinary differential equations in time on the unknown
coefficients $c_i(t)$ (see \cite{canuto06_1} for a detailed description of
spectral methods).\\ For the basis $\mathbf{u}_i$ we choose the set of
eigenvalues of the operator $\mathcal L$ formally expressed as the right hand 
side of eq. (\ref{invlapl}), with the electrical and kinematic boundary conditions of the problem.  These functions are a natural choice as elements of a
functional basis, because the features of flows at high $\ha$ are strongly
determined by the properties of this operator. For example they include 
specific features of the flow such as  laminar and turbulent
Hartmann boundary layers that develop along the channel walls
\citep{dyp09_tcfd,pdy10_jfm}. Moreover, these modes all have negative
eigenvalues, and it can be shown that to resolve the flow completely, it is
only necessary to take into account all modes with eigenvalue $\lambda$ of 
modulus below a maximum $|\lambda_{\rm max}|$, such that their total number
scales as $Re^2/\ha$ \citep{pa06_pf}, where $Re$ is the Reynolds number based 
on the large scales. Since the operator $\mathcal L$ represents
the sum of viscous and ohmic dissipation, the set of modes defined in this way
is in fact the set of \emph{least dissipative modes}.\\
The main difficulty in solving equation (\ref{invlapl}) using the least
dissipative modes lies in calculating the spectral representation of non linear
terms $G(\vel(x_i, y_i, z_i))$. We use a pseudospectral approach and calculate
these terms in real space. Therefore we need a method to reconstruct the spectral
coefficients $g_n$ of physical vector fields known at a discrete set of
points in space $\textbf{x}_i$. To this aim we first use the fact that the
eigenmodes of $\opl$ can be factorised as the product of two scalar functions
of $x$ and $y$ respectively, and a vector function of $z$.  Moreover, the
functions of $x$ and $y$ consist of Fourier modes, so the set of eigenmodes can
be enumerated by a tuple of three numbers $(n_x, n_y, n_z)$  and for every mode
we can define the vector function $\mathbf{E}_{n_x, n_y, n_z}(z)$ such that
each mode takes the form
\begin{equation} \mathbf{E}_{n_x, n_y, n_z}(z) \exp{(i k_{n_x} x + i  k_{n_y}
    y)} \,.  \end{equation}
Therefore we first calculate the two-dimensional Fast Fourier transform in the $x$ - $y$ directions.  This brings the transformed non linear terms
under the form:
\begin{equation} G(\vel(x_i, y_i, z_i)) = \sum\limits_{n_x, n_y}
 \mathbf{A}_{n_x, n_y}(z_i) \exp{(i 2\pi n_x x_i + i 2\pi n_y y_i)} \,,
\end{equation}
 where $\mathbf{A}_{n_x, n_y}$ is the complex amplitude of
Fourier mode $(2\pi n_x, 2\pi n_y)$.  Then, for every value of $(n_x, n_y)$ we
find the set of spectral coefficients $\{g_{n_x, n_y, n_z}\}$ by solving a set
of equations 
\begin{equation} 
    \sum_{n_z} g_{n_x, n_y, n_z} \textbf{E}_{n_x n_y n_z}(z_i) =
    \mathbf{A}_{n_x, n_y}(z_i) \,. \label{decFFT} \end{equation}
As the coefficients in this set of equations are constant during a single
numerical run, it is worth performing the $LU$ decomposition of the
corresponding matrix at the beginning of calculations and later use it to
efficiently find the spectral decompositions. Finally the projection onto the
subspace of solenoidal vector fields is done by neglecting the coefficients
corresponding to modes with non zero divergence.  Using the Fast Fourier
transform in $x-y$ planes imposes the distribution of discretisation points in
these planes: they have to form a regular rectangular grid. We denote its
dimensions as $N_x\times N_y$. In our simulations we also use a uniform grid in
$z$ direction of dimension $N_z$. For the set of equations (\ref{decFFT}) to
have a unique solution, the number of modes used during the spectral
decomposition has to be equal to $N_z$, and the total number of independent
modes used in the calculations is $N=N_x N_y N_z$.  The technique described
above has the advantage that the obtained spectral decomposition reproduces
exactly the physical field on the given set of discretisation points. Therefore
momentum and energy are conserved by this procedure. However the spectral 
coefficients $g_n$ obtained in this way are different from the exact ones
$\widetilde{g}_n$, that would be obtained by decomposition of
the same vector field over the space of infinite dimension spanned by all
eigenvectors of $\mathcal L$.
$|\widetilde{g}_n-g_n|$ is the so-called aliasing error.  To correct this error
we adapt the $3/2$ technique known from standard spectral methods \citep{canuto06_1}. Namely  we
perform the discrete transformation with additional number of modes $N$ larger
than the one strictly required by the system's dynamics, $N_D$ (The latter is
of the order of the attractor dimension of the dynamical system underlying the
given problem \citep{pa06_pf}).  After every evaluation of the spectral
decomposition, the coefficients corresponding to these  additional modes are
set to 0.\\
The spectral method described above was 
implemented by modifying the spectral code TARANG developed by
\cite{verma13_pramana}.
\subsection{Simulation strategy}
The bulk of our numerical simulations was based on a domain made of a cube of 
dimension $L$ divided uniformly into $N_x$,
$N_y$ and $N_z$ cells respectively in $x$, $y$ and $z$ directions.
In order to limit the dealiasing errors we always
resolve each of the Hartmann layers with at least three computational cells
in the $z$ direction. 
Our strategy to study the decay of MHD turbulence relies on four different types of simulations, all gathered in table \ref{tab:simul}.\\
The first type is inspired from the DNS of decaying MHD
turbulence in a three-dimensional periodic domain by \cite{okamoto10_jfm}: the
initial conditions consist of a isotropic, random Gaussian velocity field with $u(k) \sim
\exp{[(-k/k_p)^2]}$ where $k_p = 4 \pi / L$.  This corresponds to an energy
spectrum $E \sim k^4 \exp{[-2(k/k_p)^2]}$.  For this choice of initial velocity
field, the initial integral scale of turbulent motion is given by $l_0=\sqrt{2\pi}/k_p$.  The velocity spectrum was normalised in such a way that cell sizes
in $x$ and $y$ directions correspond to $l_K/1.4$ where $l_K = l Re^{-3/4}$ is the Kolmogorov length
scale and the Reynolds number in its definition $Re=u^\prime l/\nu$ is based on
$l$ and velocity $u^\prime=u(k=k_p)$. This strategy
allows us to calculate the most intense flow possible whilst minimising
mesh-induced numerical errors at a given mesh size, since the mesh is always
uniform.\\
To characterise the influence of the walls,  we performed additional simulations starting from exactly the same initial conditions as before, but
with periodic boundary conditions imposed in all three directions. For this set of calculations we used a traditional 
spectral code, \textsc{Turbo}, which uses Fourier modes as the functional basis and was tested and optimised for Low-$Rm$ MHD \citep{knaepen04_pf,vorobev05}.\\
% These cases were initialised with exactly the same initial conditions as the ones bounded by insulating walls at $z=\pm1$.\\
%To investigate the impact of three-dimensional effects, in particular in the later stages of the decay, we also performed a set of two-dimensional simulations.
%During these runs, we followed the evolution of two-dimensional field $\velD(x,y,t)$ through (\ref{eq:sm82}). 
%%
%We follow the evolution of the above equation with a version of \textsc{Turbo} code adopted for two dimensional setups. We initialised  $\velD$ by vertically integrating $x$ and $y$ components of the velocity field resulting from our three dimensional simulations.\\
To evaluate a possible influence of the size of the domain in the $x$ and $y$ dimensions, we also performed several simulations with a domain of dimension $2L$ in these directions. These were computationally expensive and therefore run over a shorter period of time than the simulation over a cube of size $L$.\\
% and only for the highest values of $Ha$.\\
Finally, The simple order of magnitude analysis from section 
\ref{sec:timescales} shows that under a strong magnetic field, the decay of MHD 
turbulence can occurs over a time of the order of a few times $\tau_J$, while a 
truly quasi-two-dimensional behaviour would not be expected before smaller 
scales are diffused over the height of the channel, \emph{i.e.} at much later 
times, of the order of $\tH=\ha\tJ$. Thus, if the initial integral scale $l_0$ is chosen much 
smaller than the size of the box, then turbulence will have lost practically 
all of its kinetic energy by the time the two-dimensional dynamics potentially 
becomes dominant. This would make it difficult to study the later phase of the 
decay. On the other hand, $l_0/L$ needs to be sufficiently smaller than unity for 
a significant three-dimensional phase 
of the decay to exist and so as to generate a sensible transition between three-
 and two-dimensional dynamics. To reconcile these antagonistic constraints, we 
chose $l_0/L=1/\sqrt(2 \pi)\simeq 0.4$ and also run additional 
simulations initialised with the velocity field obtained at $t=0.5\tJ,\tJ$ in 
the previous simulations, but in which the velocities are renormalised, so that 
the total energy is restored to its level at $t=0$. The flow obtained this way 
is much closer to a quasi-two-dimensional one than the simulations initialised 
 with a random field.  Comparing the evolutions of these two different types 
of initial conditions shall thus give us a good measure of the robustness of 
the features observed in the later stages of the decay to a change of 
initial conditions and to the intensity of the initial flow.
\begin{table}
    \begin{tabular}{crcccccc}
        Ha & $N_x \times N_y \times N_z$ & Boundary conditions in $z$ & $N(t=0)$ & $R\!e(t=0)$ & Energy Boost& $L_{xy}/L_z$ \\%& Designation\\
        \hline \\
        112 & $340 \times 340 \times 340$ & non slip, insulating & 8.55 & 336 & N/A & 1 \\%&\textit{Ha112}\\
        224 & $340 \times 340 \times 680$ & non slip, insulating & 34.2 & 336 & N/A & 1 \\%&\textit{Ha224}\\
        224 & $340 \times 340 \times 680$ & non slip, insulating & 39 & 261  & $t=0.5 \tJ$ & 1 \\%&\textit{Ha224}\\
        224 & $340 \times 340 \times 680$ & non slip, insulating & 43 & 229 & $t=1 \tJ$ &1 \\%&\textit{Ha224}\\
        448 & $340 \times 340 \times 1340$ & non slip, insulating& 137  & 336 & N/A &1 \\%&\textit{Ha448}\\
        448 & $340 \times 340 \times 1340$ & non slip, insulating& 161  & 246  & $t=0.5 \tJ$ &1 \\%&\textit{Ha448}\\
        448 & $340 \times 340 \times 1340$ & non slip, insulating& 170  & 225 & $t=1 \tJ$ &1 \\%&\textit{Ha448}\\
        448 & $680 \times 680 \times 1340$ & non slip, insulating& 97  & 238 & N/A &2 \\%&\textit{Ha448b}\\
        896 & $340 \times 340 \times 1340$ & non slip, insulating& 548  & 336 & N/A &1 \\%&\textit{Ha896}\\
        896 & $340 \times 340 \times 1340$ & non slip, insulating& 661  & 230  & $t=0.5 \tJ$ &1 \\%&\textit{Ha896}\\
        896 & $340 \times 340 \times 1340$ & non slip, insulating& 694  & 213 & $t=1 \tJ$ &1 \\%&\textit{Ha896}\\
        896 & $680 \times 680 \times 1340$ & non slip, insulating& 387  & 238 & N/A &2 \\%&\textit{Ha896b}\\
        112 & $340 \times 340 \times 340$ & periodic             & 8.55 & 336 & N/A &1 \\%&\textit{Ha112p}\\
        224 & $340 \times 340 \times 340$ & periodic             & 34.2 & 336 & N/A &1 \\%&\textit{Ha224p}\\
        448 & $340 \times 340 \times 340$ & periodic             & 137  & 336 & N/A &1 \\%&\textit{Ha448p}\\
        896 & $340 \times 340 \times 340$ & periodic             & 548  & 336 & N/A &1 \\%&\textit{Ha996p}\\
    \end{tabular}
\caption{Summary of parameters of calculated 3D cases. \label{tab:simul}
}
\end{table}

\section{Three dimensional phase\label{sec:3ddecay}}
\begin{figure}
\centering
\input{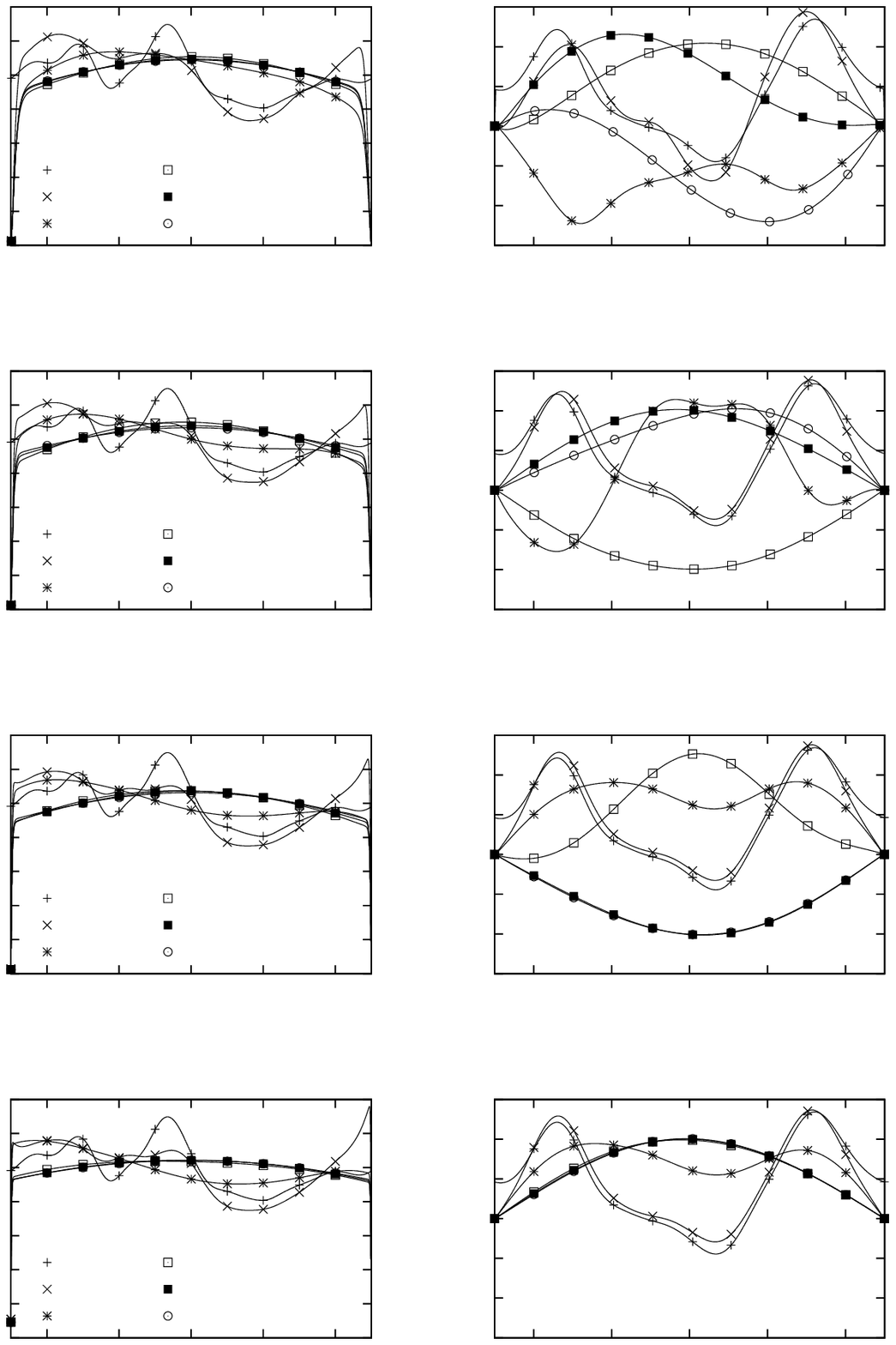}
\caption{Left column: evolution of the normalised, spatial RMS of all vertical profiles of $<\vel_\perp^2>$ over all $(x,y)$ in the domain.
% Profiles have been renormalised by their average between the walls. 
Right column: evolution of $u_z$ along a vertical line in the middle of the domain.}
\label{fig:perp}
\end{figure}

From the evolution of the total energy (figure  \ref{fig:en}), the flow 
progresses in every run through three consecutive phases: 
during the first one, the flow adjusts from the initial conditions. This phase is very short, (shorter than  $0.05 \tau_J$ in all cases). In this section, we shall characterise the phase that immediately follows, which features strongly three dimensional turbulence, a fast energy decay and lasts several Joule times. 
For the time being, the analysis shall be restricted to cubic domains of size $L$ with Hartmann walls, or with periodic boundary conditions when specified. 
This phase can be identified through the strong three-dimensionality visible in
 the spatial RMS for all $(x,y)$ of the profile along $z$ of magnitude of 
$\vel_\perp$ on fig. \ref{fig:perp}.  Significant variations along the $z$ 
direction exist until approximately 
$t\simeq 1.5-2 \tau_{2D}(l)\simeq5-10\tau_J$ for all values of $Ha$
%=56, 112$ and $2 \tau_{2D}$ at $Ha=224, 448$. 
This reflects the prominent of the contribution of the large scales to the RMS 
velocity fluctuations, and confirms that 
large scales indeed become quasi-two-dimensional in this typical time.\\ 
\subsection{Total kinetic energy}
\begin{figure}
\centering
\input{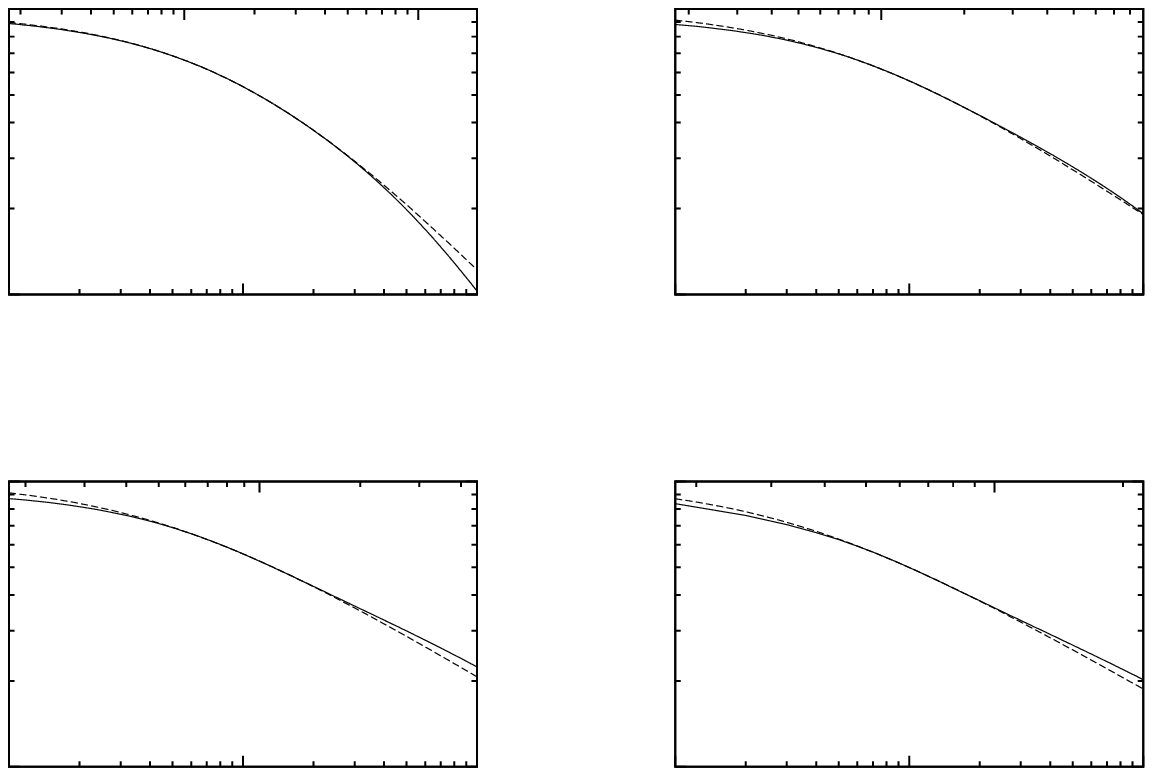}
\caption{Evolution of total kinetic energy, normalised by its value at $t=0$ (solid line). The dashed line represents the fitted law of the form $a(1+bt)^c$ over interval $[0.2, 2\tJ]$.}% For $\ha=112,224,448,896$ $a=$, $b=$ and $c=$ respectively.}
\label{fig:en}
\end{figure}

Fig. \ref{fig:en} presents the evolution of the total kinetic energy in this 
phase for different values of $\ha$. To compare this evolution to 
\cite{okamoto10_jfm}'s laws for the decay of unbounded, three-dimensional and 
initially isotropic MHD turbulence, we have fitted 
the evolution of energy to laws of the form $a(1+bt)^{c}$ in ranges from  $0.05 \tJ$ to up to $2\tJ$ ($a$, $b$ and $c$ are real constants). All values of $c$ are presented in table \ref{tab:coefc}.
\cite{okamoto10_jfm} showed that in the limit $N\rightarrow\infty$, exponent 
$c$ should be equal to $1/2$. In our cases, we obtain the best fits for 
$c=0.95$, $0.75$, $0.69$ and $0.50$ for $\ha=112$, $224$, $448$ and $896$ respectively, over $[0.05 \tJ, \tJ]$. This fit is also relatively robust to 
a variation of the fitting interval, as exponents decrease only slightly when the interval is extended to $[0.05 \tJ, 2\tJ]$.
%(\textit{and 0.79 and 0.75 for big box with $\ha=448$ and 896}). 
From this, we infer that the decay of turbulence between walls is in this line 
with these authors' prediction over a duration of about $\tJ$.
Values of $c$ are however slightly higher for cases with periodic boundary 
conditions over this interval, which suggests that the influence of the walls 
is present but moderate for $t\lesssim\tJ$. This 
phase corresponds to roughly 25-50\% of the time interval where we identified 
strong three-dimensionality in the profiles of RMS velocity fluctuations 
(figure \ref{fig:perp}). 
From times $t>\tJ$, by contrast, the kinetic energy tends to decay at a slightly slower rate than predicted by \cite{okamoto10_jfm} in all cases. This is an indication that some of the turbulent structures interact with the walls, as they become stretched vertically under the effect of diffusion by the Lorentz force. The energy of such structures is dissipated partly by the action of eddy currents recirculating in the Hartmann layers. This Hartmann friction mechanism is typically $Ha$ times slower than Joule dissipation, which would be the unique electromagnetic dissipation mechanism if these structures were not in contact with the wall. This explains that the decay of energy slows down for $t\gtrsim\tJ$.\\ 
All cases where the energy was boosted during the three-dimensional phase
exhibit a similar behaviour to cases where the energy was left to decay from
the start. Nevertheless, for all of them, the value of the $c$ coefficient 
fitted over intervals of $\tJ$ or more is lower than for their counterpart 
without energy boost. 
Furthermore, the later the energy is boosted, the lower the value of $c$.
 %($c=0.49$ at $Ha=224$ and $c=0.4$ at $Ha=448$ for an energy boost at $\tJ/2$). 
This slower decay reflects the influence of the anisotropy of the boundary 
conditions: at the time of the energy boost, the flow recovers the same energy 
as the initial one but conserves the anisotropy that has developed during the initial decay, before the energy was boosted.
Consequently, vortices are more elongated, interaction with the walls is
more significant and the slower friction in the Hartmann layers represents a more important fraction of the dissipation.
\begin{table}
\centering
\begin{tabular}{llcccc|cccc}
&& \multicolumn{4}{c|}{E fitted from $0.05\tJ$ to  } & \multicolumn{4}{c}{$l_z$ fitted from $0.05\tJ$ to} \\
\hline
$\ha$ & Remarks&$2\tJ$ &$\tJ$ & $0.5\tJ$ & $0.2\tJ$&
$2\tJ$& $\tJ$ & $0.5\tJ$ & $0.2\tJ$ \\
\hline
112 &  & -0.91 & -0.95 & -0.97 & -0.86 & 0.18 & 0.21 & 0.24 & 0.26 \\
224 &  & -0.65 & -0.75 & -0.82 & -0.71 & 0.26 & 0.33 & 0.37 & 0.42 \\
448 &  & -0.59 & -0.69 & -0.42 & -0.83 & 0.27 & 0.34 & 0.38 & 0.44 \\
896 &  & -0.51 & -0.51 & -0.41 & -0.34 & 0.26 & 0.34 & 0.37 & 0.44 \\
224 &E boosted at $0.5\tJ$& -0.54 & -0.58 & -0.63 & -0.68 & 0.20 & 0.25 & 0.28 & 0.29 \\
448 &E boosted at $0.5\tJ$&  -0.46 & -0.53 & -0.73 & unstable & 0.20 & 0.26 & 0.29 & 0.40 \\
896 &E boosted at $0.5\tJ$& -0.45 & -0.56 & unstable & unstable &       0.21 & 0.29 & unstable & unstable\\
224 &E boosted at $1\tJ$&  -0.50 & -0.51 & -0.52 & -0.56 & 0.15 & 0.20 & 0.22 & 0.22\\
448 &E boosted at $1\tJ$&  -0.43 & -0.48 & -0.70 & unstable & 0.18	& 0.23 & 0.25 & 0.39\\
896 &E boosted at  $1\tJ$&  -0.43 & -0.52 & unstable & unstable & 0.19	& 0.24 & 0.36 & unstable\\
112 &periodic BC&  -0.76 & -0.86 & -0.96 & -1.05 & 0.33	& 0.37 & 0.40 & 0.42\\
224 &periodic BC&  -0.56 & -0.69 & -0.79 & -0.74 & 0.40	& 0.50 & 0.55 & 0.59\\
448 &periodic BC&  -0.52 & -0.65 & -0.82 & -1.02 & 0.42	& 0.50 & 0.56 & 0.64 \\
896 &periodic BC&  -0.47 & -0.58 & -0.66 & -0.89 & 0.43	& 0.51 & 0.57 & 0.65 \\
448 &Domain $2L\times2L\times h$ & -0.91 & -1.04 & -1.14 & -1.22 & 0.38 & 0.49 & 0.56 & 0.66 \\
896 &Domain $2L\times2L\times h$ & -0.87 & -1.03 & -1.29 & -1.68& 0.37 & 0.46 & 0.45 & 0.38 \\
\hline
\end{tabular}
\caption{Fitted values for constant $c$ in \cite{okamoto10_jfm}'s laws of the form $a(bt+1)^c$ for the decay of total kinetic energy and the growth of integral scale $l_z$. Cases where the formal error of the least squares method was 
greater than 20\% are marked as "unstable".\label{tab:coefc}}
\end{table}
%
%\begin{table}
%\begin{tabular}{lccc}
%run & $c_1$ & $c_{0.5}$ & $c_{0.2}$ \\
%\hline \\
%$\ha=112$ & 0.21 & 0.24 & 0.28\\
%$\ha=224$ & 0.33 & 0.36 & 0.44\\
%$\ha=448$ & 0.34 & 0.36 & 0.41\\
%$\ha=896$ & 0.340 & 0.37 & 0.62\\
%$\ha=112$ periodic & 0.37 & 0.40 & 0.42\\
%$\ha=224$ periodic & 0.50 & 0.55 & 0.60\\
%$\ha=448$ periodic & 0.51 & 0.55 & 0.68\\
%$\ha=896$ periodic & 0.54 & 0.57 & 0.65\\
%$\ha=224$ boosted & 0.20 & 0.22 & 0.22\\
%$\ha=448$ boosted & 0.23 & 0.26 & 0.39\\
%$\ha=896$ boosted & 0.23 & 0.25 & 0.34\\
%\end{tabular}
%\caption{Fitted coefficients for $l_z$}
%\end{table}
%
\subsection{Dissipation}
\begin{figure}
\centering
\input{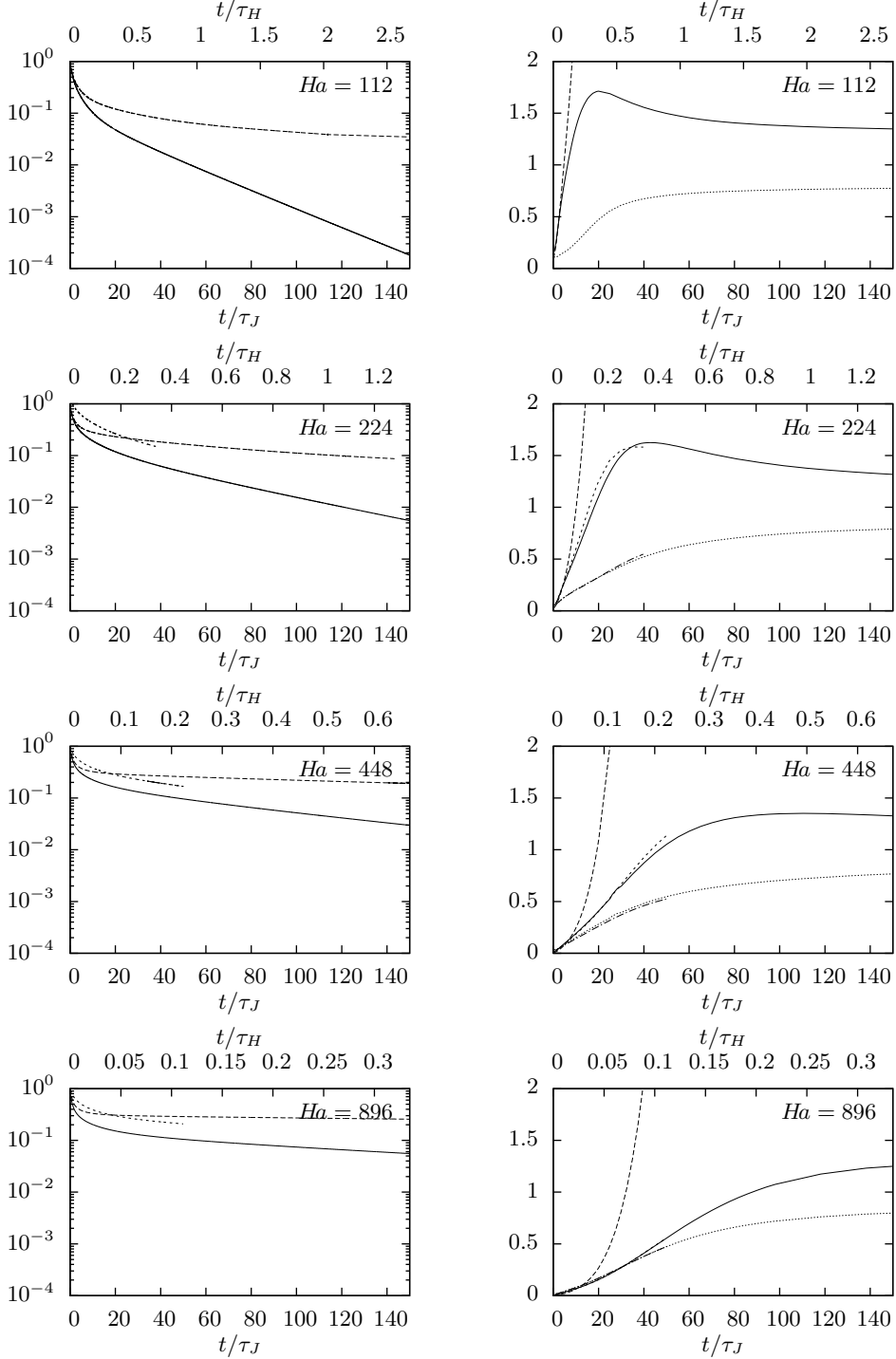}
\caption{Left column: evolution of the kinetic energy in cases with walls (solid lines), periodic boundary conditions (dashed line). Right column: ratio of viscous to Joule dissipation in cases with walls (solid line) and periodic boundary conditions (dashed line). The dotted line represent the fraction of energy which is dissipated in Hartmann layers in cases with wall. On all graphs, the shorter curves represent the case with walls where energy was boosted at $t=0.5\tJ$ 
(Short dashed line: Joule to viscous dissipation, short dash-dotted line: fraction of dissipation in the Hartmann layers) .
\label{fig:diss}}
\end{figure}

Initially, the kinetic energy is mainly dissipated ohmically and the initial
ratio of viscous to Joule dissipation scales as $\sim 1/\ha$.  As the flow
becomes more two-dimensional, both viscous and Joule dissipations diminish in 
the bulk of the flow (between the Hartmann layers). Conversely, dissipation 
in the Hartmann boundary layers increases as more and more structures interact
with the walls. Therefore,  the main contribution from the total dissipation ends up coming from the Hartmann layers (see fig. \ref{fig:diss}).
For $\ha=112$, $224$, $448$ and $896$ dissipation in the layers becomes larger
than in the bulk at $t=25 \tJ$, $37.6 \tJ$, $45.1 \tJ$ and $54
\tJ$ respectively. However in the Hartmann layers the Joule dissipation is
nearly the same as viscous dissipation. Therefore the global ratio of viscous
to Joule dissipation increases with time. The value of this ratio becomes 
larger than unity for $t=6.3 \tJ$, $19 \tJ$, $50.1 \tJ$ and $88 \tJ$ (or equivalently at $t=0.056 \tau_H$, $0.084 \tau_H$, $0.11 \tau_H$ and $0.2 \tau_H$)  for $\ha=112$, $224$, $448$ and $896$
respectively. In all cases, the ratio $\epsilon_\nu/\epsilon_J$ tends to an asymptotic value of $\sim 1.3$ after a time of the order of $\tH$. From this perspective, the dissipation behaves as in a three-dimensional flow during a period of time that is longer than the timescale of three-dimensional Joule dissipation 
$\tJ$, but shorter than that of two-dimensional effects $\tH$. 
This intermediate time scale is an indication that although large scales become 
two-dimensional over a time of the order of $\tau_{2D}(l)$, smaller scales 
remain three-dimensional during a significantly longer period of time. After 
$t>1.5-2\tau_{2D}$, the contribution of the large scales to the total 
dissipation comes mostly from the Hartmann layer, where it is weak, whereas the 
contribution of the small scales consists of stronger Joule dissipation in the 
bulk. Joule dissipation therefore remains higher than viscous dissipation 
until this contribution has significantly reduced, which occurs on a timescale 
of at most $\tH$ (see section \ref{sec:timescales}).
%On the other hand, it does not exhibit the behaviour of quasi-two dimensional flows before a time of the order of $\tau_H$. 
%
\subsection{Integral lengthscales}
\begin{figure}
\centering
%\caption{Evolution of integral lengthscale $l_z$.}
%\label{fig:lzlxy}
%\end{figure}
%
%\begin{figure}
%\centering
\input{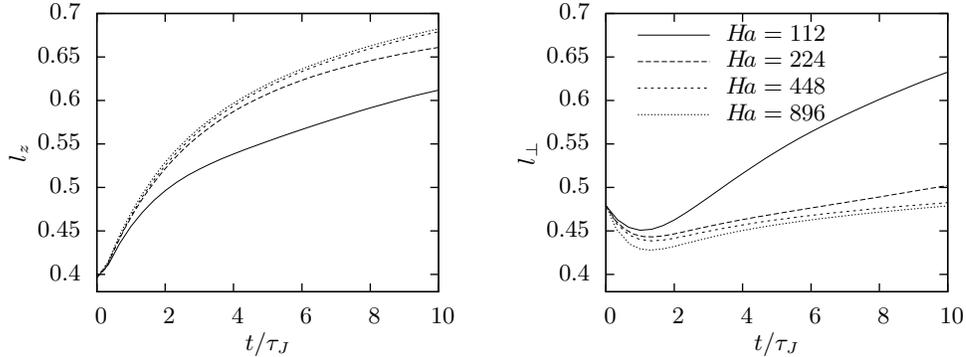}
\caption{Evolution of integral lengthscales $l_z$  (left) and $l_\perp$ (right) in the presence of Hartmann walls.}
\label{fig:lzlxy}
\end{figure}

% At
%this stage the integral lengthscale $l_z$ can be also fitted with formula
%$a(1+bt)^{-c}$ with $c\approx0.15$
Figure \ref{fig:lzlxy}  (top) shows the initial evolution of the integral lengthscale in the $z$ direction $l_z$, and in the direction orthogonal to the magnetic field $l_\perp$. These are respectively defined as:
\begin{align}
l_z &= \frac{\int\int{\!u_z(x,y,z) u_z(x,y,z+z')\,\mathrm{d}V} dz'}{\int\!u_z^2(x,y,z)\,\mathrm{d}V} \\
l_\perp &= \frac{1}{2} \left( \frac{\int\int{\!u_x(x,y,z)\cdot u_x(x+x',y,z)\,\mathrm{d}V} dx'}{\int\!u_x^2(x,y,z)\,\mathrm{d}V} \right. \nonumber\\
&+ \left. \frac{\int\int{\!u_y(x,y,z)\cdot u_y(x,y+y',z)\,\mathrm{d}V} dy'}{\int\!u_y^2(x,y,z)\,\mathrm{d}V} \right). \label{eq:lzuz}
\end{align}
The initial growth of $l_z$ can be again fitted with the formula $a(1+bt)^{c}$. Except for $t<0.2\tJ$, the fitted value of $c$ is significantly smaller than 
\cite{okamoto10_jfm}'s theoretical value of 0.5. It is also strongly dependent 
on the fitting interval for $Ha=224$, $448$ and $896$. For $Ha=112$ the integral lengthscale $l_z$ grows even more slowly with a fitted value down to
$c\approx0.18$ over $[0.05, 2\tJ]$. This behaviour indicates a very strong 
influence of the walls on the growth of $l_z$ from the outset of the decay.
It is somewhat remarkable that despite this early influence of the walls on 
$l_z$,  the energy decay shows little influence of the walls during as long as 
 $\tJ$ -- $2 \tJ$.
%\textit{$c=0.4$ and $c=0.36$ for big boxes with $\ha=448$ and 896}
During this initial stage $l_z$ also grows faster for larger values of $Ha$, 
with indication that at $Ha=224$, it is already close to its asymptotic 
behaviour (in the sense of large $\ha$). For $t\gtrsim 1.5\tJ$, the growth of 
$l_z$ slows significantly. As for the energy, this is due to the increasingly wide range of 
scales at which structures reach the walls during the two-dimensionalisation 
process, and whose growth in thus impeded in the $z$ direction.\\
With periodic boundary conditions, the fitted value of exponent $c$ is higher 
than with Hartmann walls. It is close to the theoretical value of 0.5 at high 
$\ha$ for $t<\tJ$, and decreases thereafter. Unlike in cases with walls, this 
behaviour is quite insensitive to 
the fitting interval for $t<\tJ$, which confirms the validity to 
\cite{okamoto10_jfm}'s law at high $\ha$ over at most one Joule time.
% ($c=0.33$ for $\ha=112$, $c=0.4$ for $\ha=224$, $c=0.42$ for $\ha=448$ and $c=0.43$ for $\ha=848$). 
The validity of this law also indicates that the parameter $l_\perp(t=0)/L$ 
was chosen sufficiently small to observe the main features of three-dimensional
 unbounded turbulence in a periodic domain. However, the fact that this 
exponent is higher with periodic boundary conditions than with walls suggests 
that eddy currents circulating between the Hartmann layer and the bulk (when 
walls are present) strongly increase the influence of the boundaries compared 
to the periodic case where this effect is absent.\\ 
%Interestingly, even though the evolution of $l_z$ is strongly
%affected by the presence of the walls from the outset, the energy shows a 
%slower decay after a period of $\tJ$ -- $2 \tJ$, only.\\
%This is confirmed by the fact that fitting a law of the form $a(1+bt)^c$ 
%to the evolution of $l_z$ over much shorter intervals ($[0.05\tJ-0.15\tJ]$) yields values of $c$ that are much closer to the theoretical value of 
%0.5 for unbounded turbulence.\\
%At later times ($t\gtrsim \tau_H$), $l_z$ increases at a slower rate.\\
The integral lengthscale $l_\perp$ grows very slowly during the decay. This is 
consistent with the prediction of \cite{okamoto10_jfm} of a growth as 
$(t/\tJ)^{1/7}$. However such a small exponent is difficult to quantify on a 
timescale of the order of $\tJ$, where it is expected to be valid. 
Furthermore, $l_\perp$ evolves slowly all the way through our calculations, 
with no clear evidence of different behaviour when the flow is close to 
two-dimensional than when it is three-dimensional.
%
% Alternatively use this for the bibliography :
%
%\begin{thebibliography}{1}
%
%\bibitem{bib:momo1965} 
%A. Momo, B. Mimi, and C. Mama. Experimental study of blibli. \textit{Journal of Blibli} \textbf{15}: 43--62, 1965.
%
%\bibitem{bib:toto2002} 
%A. Toto, B. Titi, Tutu and C. Tutu. Effect of blibli on blublu. \textit{Journal of Blabla} \textbf{468}: 77--105, 2002.
%
%\end{thebibliography}

\section{Quasi-two dimensional phase \label{sec:2ddecay}}
We shall now describe the later stage of the flow evolution where 
it approaches a quasi-two dimensional behaviour, and characterise this asymptotic regime. 
%5third phase which features two dimensional and quasi exponential decay of the energy. 
\subsection{Velocity profiles \label{sec:velo}}
\begin{figure}
\centering
\input{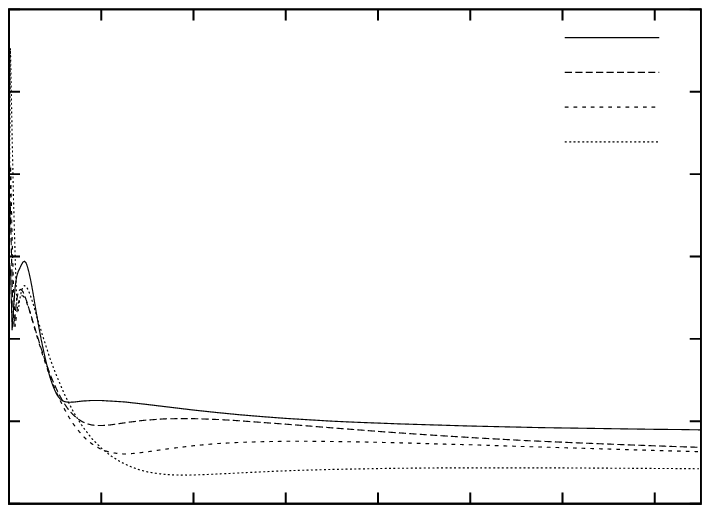}
\caption{Evolution of the maximum of $<\vel^2_\perp>_{xy}/\overline{<\vel^2_\perp>_{xy}}$ 
(as represented on figure \ref{fig:perp}).}
\label{fig:maxu}
\end{figure}

When this stage is reached, the spatial RMS over all $(x, y)$ of the profile along $z$ of $\vel_\perp$ have already been considerably smoothed out during the three-dimensional phase of the decay. 
%5 converges to the
%\textit{barrel like} shapes.
 In every case, we were able to identify a time $\tQ$, from which the profile 
starts to flatten monotonically without qualitatively changing shape. This type 
of decay would be expected from a flow governed by mostly two-dimensional 
dynamics. $\tQ$ was defined as the time at which the maximum value in the  
velocity profile starts decreasing monotonically. It was found at 
$19.3 \tJ (=0.34 \tH)$, 
$37.6 \tJ (=0.336 \tH)$, $65.2 \tJ (=0.29 \tH)$ and $98 \tJ (=0.22 \tH)$ for 
$\ha=112$, 224, 448 and 896 respectively.  The fact that it obeys a timescale 
of about $0.3\tH$ that is commensurate with the two-dimensional timescale 
$\tau_H$ indicates that 
the large scales only acquire a two-dimensional dynamics once a 
significant part of the spectrum is close to being two-dimensional (since 
two-dimensionalisation of the whole spectrum is expected to occur over a period 
of approximately $\tau_H$). For $t>\tQ$, the evolution of the shape of the 
profiles is practically unaffected by smaller scales that still 
retain a three-dimensional behaviour at this stage of the decay.\\
% The contribution form the remaining three-dimensional scales accounts for a
Remarkably, in none of the cases, did we find that the profile was quasi-two-dimensional after $t\gtrsim \tH$ (although such very long times could not be reached for $\ha=448$). Instead, all profiles seem to have reached a barrel-like shape after $t\sim 50 \tJ$, which evolves only very slowly after this time. 
Also the shape of the profile is flatter for the larger values of $\ha$. 
%This indicates that the large scales are quasi two dimensional at this stage. 
This shape was first theorised by \cite{psm00_jfm} and numerically observed by \cite{muck00_jfm}. It stems from eddy currents recirculating between the Hartmann layers
 and the core. These currents are driven along the axis of columnar vortices. Their leak into the core drives differential rotation between horizontal planes of the vortex leading to the barrel-shaped profile (see also \cite{p12_epl}).
 Since the effect is driven by currents recirculating between the bulk and the 
Hartmann layer, which scales as $\ha^{-1}$, it is less marked at high $\ha$.\\
The intensity of the barrel effect can be measured through the relative value 
of the maximum in the profile. The evolution of this quantity is shown on figure \ref{fig:maxu}. First, this graph confirms quantitatively that the barrel 
effect is less pronounced at higher $\ha$.
%which is consistent with \cite{psm00}' theory. 
Second, the graph also confirms the invariance of the barrel shape of the large scales beyond $t\sim50\tJ$. This appears to be verified over at least $100\tJ$, 
although the total energy drops by a factor of up to a couple of orders of magnitude during this interval, depending on the value of $\ha$. This remarkable feature shows that even at high $\ha$, a form of three-dimensionality 
subsists a large times, even in the large scales.\\
%stems from the electric nature of the barrel effects: since it is
%driven by currents recirculating between the bulk and the Hartmann layer, 
%the current density in the core, it is proportional to the ratio of their thicknesses, the Hartmann number.
Finally, we verified that cases where the 
energy was boosted at $t=0.5\tJ$ and $t=\tJ$ exhibit the same behaviour, which 
indicates robustness to initial conditions of this scenario.
\subsection{Energy}
In all calculated cases (with and without energy boost), the evolution of the total kinetic energy in the late phase can be described as 
an exponential decay with slowly changing timescale. However even at the latest time in our calculations this timescale does not seem to converge towards a clear asymptotic value. 
%The only exception is the run with $\ha=56$ which shows indications of a pure exponential decay with characteristic timescale $2.28 Ha$ for $t>120 \tJ=2.1\tH$.
 This indicates that even for $t\gtrsim \tH$, 
when all structures could theoretically be expected to have been diffused 
across the channel height, the decay is still not entirely dominated by 
dissipation in the Hartmann layers.
% \textcolor{red}{Kacper, would it be possible to compare the evolution of the  $E$ with that  of $E(t=t2D)+\int_{t2D}^t -\nu\|\nabla \mathbf u\|^2 -E/t_H dt$: this quantity would in principle cover of 2D decay mechanisms including viscous friction. Any discrepancy between it and $E$ would be a definite signature of 3D effects.}  
The origin of the extra dissipation shall be determined by examining different types of dissipation in section \ref{sec:2d_dissip}.\\

\begin{figure}
\centering
\input{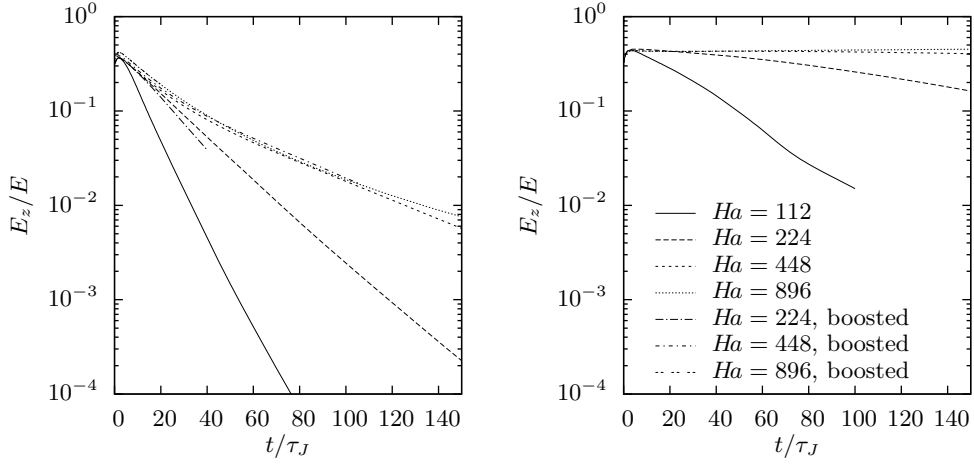}
\caption{Evolution of $E_z/E$ for cases with walls (left) and with periodic 
boundaries (right).\label{fig:ez}}
\end{figure}

% in the evolution the 3D structures contribute significantly to the dissipation.\\ 
The energy associated to the $z$ component of the velocity is represented on 
figure \ref{fig:ez}. For all cases with Hartmann walls, $E_z/E$ tends rapidly 
to zero, which is consistent with the experimental findings of 
\cite{kolesnikov1974_fd} for turbulence in a duct. By contrast, experiments 
where turbulence was kept far from Hartmann walls 
\citep{alemany79} and simulations with periodic boundary conditions rather than 
walls (\cite{schumann76,burattini10_jfm}) show a transfer of 
energy to the velocity component along $\mathbf B$, which results in a maximum 
in $E_z/E$, followed by a much slower decay than in the case with walls. This 
supports the explanation put forward by \cite{burattini10_jfm} who hypothesised 
that the difference between these two scenarii was due to the presence of the 
Hartmann walls.

%These results are also consistent with \cite{dav97}'s theory who attributes 
%the transfer of energy to $E_z$ to the constraint of conserving angular 
%momentum aligned with $\mathbf B$ whilst $E$ decays. When Hartmann walls are 
%present, 
%this constraint is indeed violated because of the friction in the Hartmann layers and so all components of the angular momentum decay, as well as $E_z$.\\
%
\begin{figure}
\centering
\input{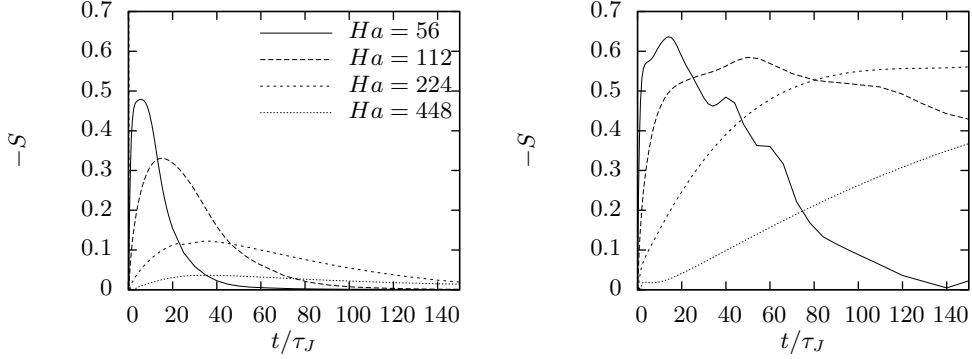}
\caption{Evolution of Skewness S for cases with walls (left) and with periodic   
boundaries (right).\label{fig:skew}}
\end{figure}

The question of how much transport along $\mathbf B$ remains asymptotically 
can be further analysed through the evolution of the skewness coefficient 
\begin{equation}
S=\frac1{35}\left(\frac{15}{\hat\epsilon}\right)^{3/2}\hat\Gamma,
\end{equation}
where
\begin{eqnarray}
\hat\Gamma=\sum_{\mathbf k} 2k^2 \hat{\mathbf q}(\mathbf k)\cdot\hat{\mathbf u}^*(\mathbf k),\\
\hat\epsilon=\sum_{\mathbf k} 2k^2 \hat{\mathbf u}(\mathbf k)\cdot\hat{\mathbf u}^*(\mathbf k),
\end{eqnarray}
which \cite{schumann76}, found to remain constant at large times. 
This author interpreted this behaviour as an evidence that transport of $E_z$ 
became important at large times. Our computed evolution of $S$ is shown on 
figure \ref{fig:skew}. 
In the domain with periodic boundary conditions, we recover \cite{schumann76}'s 
findings at high $Ha$ that $S$ seems to converge to a constant asymptotic 
value. However, as in this authors' work, we were not able to compute 
$S$ for $t>\tH$ at the highest value of $Ha$, so the two-dimensionalisation 
process may not be entirely complete at the end of this particular calculation. 
This leaves room for the possibility that $S$ may in fact evolve on a timescale 
significantly longer than the duration of our calculations in this case. 
Simulations with Hartmann walls, by contrast, exhibit a fast decay of $S$ and 
it is readily visible that $S\rightarrow0$ in all calculated cases.
Together with the fast decay of $E_z/E$, this behaviour confirms that 
the presence of walls results in a very strong suppression of transport 
along $\mathbf B$, in contrast to the phenomenology of flows with periodic 
boundary conditions. 

Note that in the two-dimensional phase, both $E$ (see fig.~\ref{fig:diss}) and 
$E_z/E$ 
appear to decay faster at lower values of $\ha$ in the case with walls. It 
should however be pointed out that in this phase, turbulence decays over a 
timescale of $\tH=\ha\tJ$, which appears slower at higher $\ha$ in units of 
time normalised by $\tJ$ when it is in fact faster in dimensional time units.\\
%Nevertheless, with periodic boundary conditions, $E_z/E$ does not seems to 
%converge to 0, when in the limit $\ha\rightarrow\infty$, $E_z/E$ seems to tend 
%to a value ever closer to the value of 1/3 predicted by the linear theory 
%\citep{moffatt67}. Since for $\ha=0$ with periodic boundary conditions, the 
%ratio $E_z/E$ is not expected to depart significantly from 1/3 throughout the 
%decay, this suggests that the asymptotic value of this ratio may not vary monotonously with $\ha$.\\ 
%This particularity was not noticed in previous studies 
%probably because it is only visible at very long times and for rather high values of the initial interaction parameter (the decay at $\ha=224$, for instance is very slow). 
Finally, in simulations where the total energy in the flow was artificially 
boosted at $t=0.5\tJ$ (and, we verified, $t=\tJ$) both $E$ and $E_z/E$ exhibit 
a similar behaviour to the case without energy boost, which confirms the
robustness of this phenomenology to initial conditions.
\subsection{Dissipation \label{sec:2d_dissip}}
\begin{figure}
\centering
\input{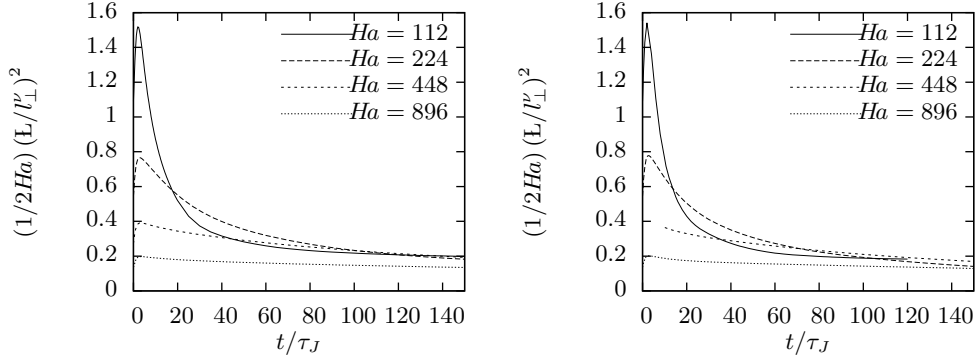}
\caption{Ratio of effective dissipation by two-dimensional viscous friction to Hartmann friction: case with Hartmann wall (left) and case with periodic boundary
conditions (right). With periodic conditions, this quantity represents the normalised two-dimensional dissipation only, as Hartmann friction is absent.}
\label{fig:2d_diss}
\end{figure}

Asymptotically, the ratio of viscous to ohmic dissipations 
$\epsilon_\nu/\epsilon_J$ tends to a value of approximately $1.3$ for all values of 
$\ha$ (see fig. \ref{fig:diss}). For all cases
except $\ha=448$ we observe a maximum in the temporal evolution of this ratio 
at times $0.35 \tH, 0.38 \tH, 0.5 \tH$ for $\ha=112$, 224 and 448 
respectively. The maximum becomes less pronounced with increasing $\ha$, with 
the ratio of maximal value of $\epsilon_\nu/\epsilon_J$ to
its asymptotic value decreasing from 1.28 at $\ha=112$ to 1.04 at $\ha=448$.
We presume that for $\ha=896$, this maximum is still present for 
$t\gtrsim 0.5 \tH$, although even less prominent.   
This behaviour is very different from that observed in simulations with 
periodic boundary conditions, where $\epsilon_\nu/\epsilon_J$ increases 
indefinitely. In both 
cases the sharp initial increase of this ratio is due to the 
two-dimensionalisation of smaller and smaller structures. Indeed, from the 
right hand side of (\ref{invlapl}), the ratio of 
dissipations for a structure of wavenumber $(k_\perp,\kappa_z)$ is 
$\epsilon_\nu(k_\perp,\kappa_z)/\epsilon_J(k_\perp,\kappa_z)=\ha^{-2}(1+(k_\perp/\kappa_z)^2)$. While initially $k_\perp/\kappa_z\sim 1$,  $k_\perp/\kappa_z$
becomes very large for nearly two-dimensional structures. As time progresses, 
this becomes true for increasingly large values of $k_\perp$, and so the ratio 
of total viscous to Joule dissipations $\epsilon_\nu/\epsilon_J$ increases. 
With periodic boundary conditions, strictly two-dimensional structures ($\kappa_z=0$) can exist and so this ratio is unbounded. 
In the presence of Hartmann walls, on the other hand, quasi-two-dimensional 
structures generate little 
dissipation in the bulk. Most of their dissipation comes from 
the Hartmann layers, where viscous and Joule dissipation are locally of the 
same order. %As time progresses, two-dimensionality is achieved in increasingly small structures and 
This explains that the ratio $\epsilon_\nu/\epsilon_J$ converges to a value of 
the order of unity.\\
 The presence of the maximum in the evolution of $\epsilon_\nu/\epsilon_J$ at 
lower values of $\ha$ may be attributed to structures that are too small for 
magnetic diffusion to stretch them up to the walls, i.e. for which at $t=0$, 
$l_z^N(k_\perp)=k_\perp^{-1}N(k_\perp)^{1/2}<L$. These initially generate a 
large contribution to the ratio $\epsilon_\nu/\epsilon_J$ over a time scale of 
the order of $\tD(k_\perp)$, which is longer than the 
two-dimensionalisation time of the large scales $\tD(l_0)$ but shorter than 
the decay time of quasi-two-dimensional structures $\tH$. For $t\sim \tH$, 
these structures have lost most of their energy and the surviving structures 
extend across the whole channel. As $\ha$ increases, structures that remain 
three-dimensional during their entire life span are confined to a 
region of the spectrum of higher and higher lengthscale and their contribution 
to the total dissipation progressively vanishes. This explains that the 
maximum is less pronounced at higher values of $\ha$, and also that it take 
place at later times.\\
More details on the asymptotic state can be obtained by inspecting the ratio of the total dissipation coming from the Hartmann layers to that coming from the bulk 
(figure \ref{fig:diss}). Asymptotically, the dissipation comes dominantly from 
Hartmann layers because the flow becomes close to quasi-two dimensional at all 
values of $\ha$ investigated here. 
However, approximately 20\% of the dissipation still takes place in the bulk 
asymptotically. It is tempting to check whether the residual 
three-dimensionality due to the Barrel effect noted on the velocity profiles 
 (section \ref{sec:velo}) is responsible for this extra dissipation: since it 
is driven by currents in the Hartmann layers that recirculate in the 
bulk, Joule dissipation must be associated to it. Current conservation implies 
that the current densities in the bulk $J^C$ and the Hartmann layers $J^H$ must 
satisfy $J^H/J^C\sim \ha$ and so the contributions to Joule dissipation from 
the core and the Hartmann layers must be in a ratio of $\epsilon_J^H/\epsilon_J^C\sim \ha$. This scaling is insufficient to explain the relatively high level 
of dissipation observed in the bulk: this discards the barrel effect as the 
residual source of dissipation. 
To find out the origin of this extra dissipation, we calculated non-dimensional
quantity $2\ha^{-1}(L/l_\perp^\nu)^2$, which, from (\ref{eq:2d_energy}),
represents the actual ratio of quasi-two-dimensional viscous friction to
Hartmann friction. It turns out that this ratio converges towards a constant
value around 0.2 regardless of the case considered (see figure \ref{fig:2d_diss}). This implies that
extra dissipation in the bulk results from viscous friction incurred by
gradients of the $x$ and $y$ components of the velocity in the $x$ and $y$
directions.
%\, which represents up to half of the dissipation in the Hartmann
%layers (hence about a third of the total dissipation). 
This extra dissipation also explains that the energy does not decay 
exponentially even for $t\gtrsim\tH$, as would be expected if Hartmann 
friction was the exclusive dissipation mechanisms.
 Interestingly, the ratio $2\ha^{-1}(L/l_\perp^\nu)^2$ behaves in roughly the same way whether Hartmann 
walls are present or not. Even though Hartmann friction is absent when 
boundary conditions are periodic in all three spatial directions, this 
ratio still gives a normalised measure of the level of two-dimensional 
viscous dissipation, which appears to be the same with and without Hartmann walls.
% The importance of 
%viscous friction is visible on the 
%evolution of $\ha/Re$ on figure \ref{fig:hare}: in the later stages of the 
%decay, the viscous friction in these planes is less than an order of magnitude 
%less than inertia ($Re<10$) and two-dimensional inertia is itself of the same 
%of order as friction in the Hartmann layer.
Furthermore, 
% the electric current is recirculated between the bulk of the flow and the layers. Asymptotically 80\% of energy is dissipated in the Hartmann layers. 
our simulations show that in the Hartmann layers, Joule and viscous dissipations
 are asymptotically the same up to few percent. Therefore viscous dissipation 
due to horizontal velocity gradients also explains that the ratio 
$\epsilon_\nu/\epsilon_J$ remains greater than unity even for $t\gtrsim \tH$, 
when two-dimensionalisation is expected be complete over the whole spectrum of 
lengthscales. 
\begin{figure}
\centering
\input{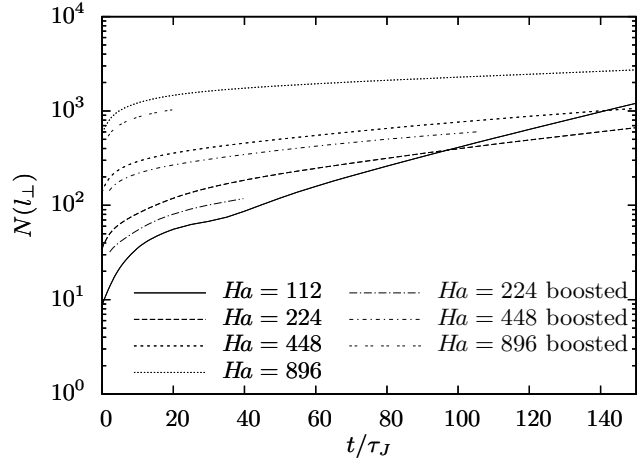}
\caption{Evolution of interaction parameter $N=B l_\perp / u$, }
\label{fig:N}
\end{figure}
%
%\begin{figure}
%\centering
%\input{hare}
%\caption{Evolution of $Ha/R\!e$.}
%\label{fig:hare}
%\end{figure}
%
%the excess of total dissipation in the bulk must be viscous over ohmic dissipation
%comes from the bulk of the flow.  Here most of the viscous dissipation has an
%origin in gradients of $x$ and $y$ components of the velocity in $x$ and $y$
%directions.
%
\subsection{Integral lengthscales} 
\begin{figure} \centering
    \input{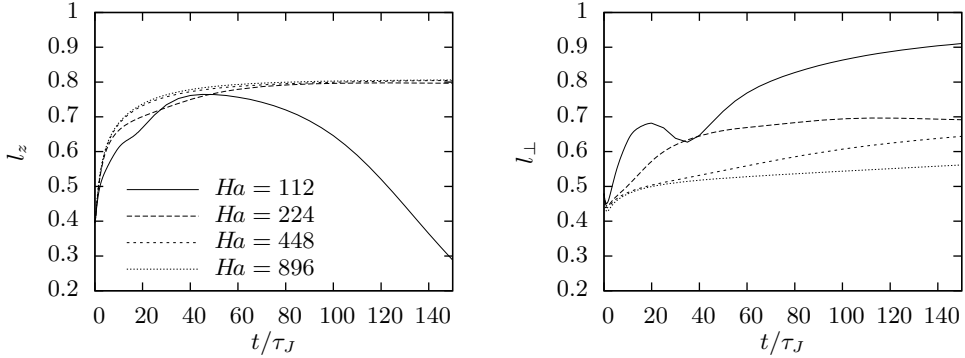}
    \caption{Evolution of integral lengthscales $l_z$ (left) and $l_\perp$ (right) for the simulation with walls.} \label{fig:lz_f}
\end{figure}

Figure \ref{fig:lz_f} shows the evolution of $l_z$ up to the later stages 
where $t>\tH$. For large values of $\ha$, $l_z$ tends asymptotically to a value 
of approximately $0.8$. 
% This value is again the result of the \textit{barrel} effect. 
By contrast, in simulations with periodic boundary conditions 
%instead of walls, the effect does note appear and 
$l_z$ asymptotically tends to 1. At high $\ha$ the Hartmann layers are laminar 
and not affected by inertia present in the core. The classical theory for these 
layers then implies that the velocity normal to the wall in the layer is 
$\mathcal O(\ha^{-1})$ and this imposes a region of very low values of $u_z$ 
near the walls. By contrast, residual velocity may exist in the core and so 
the z-component of the velocity cannot be correlated over the entire width of 
the channel (see profiles of vertical velocity on figure \ref{fig:perp}). This 
effect is absent with periodic boundary conditions where a constant 
through-flow can exist in the z-direction, which allows values of $l_z$ close to 1.\\
The convergence of $l_z$ for cases bounded by walls becomes less smooth for 
lower values of $\ha$. For $\ha=112$, $l_z$ shows no signs of convergence to a finite value. It starts decreasing around $t\sim 45 \tJ$ and still decreases at $t\simeq 3\tH$.  This behaviour is caused by Ekman pumping in large quasi-two dimensional vortices. This effect tends to produce vertical profiles of
$u_z$ that are antisymmetric with respect to the midplane. According to 
(\ref{eq:lzuz}), reduces the value of $l_z$ as its definition is based on 
$u_z(z)$ (see fig. \ref{fig:perp}). The intensity of Ekman pumping is driven 
by inertia but damped by the Lorentz force: \cite{psm00_jfm} showed 
that it scaled as $u_z/u_\perp\sim Ha^{-2}N^{-1}$. This explains that this 
effect is only noticeable at the lowest value of $\ha$. This also suggests 
that ultimately, since the interaction parameter $N(t)$ diverges asymptotically 
(see figure \ref{fig:N}), Ekman pumping should progressively 
disappear and $l_z$ may increase again. The evolution of $l_z$ on figure 
\ref{fig:lz_f} however implies that this may only take place after a very 
long time, beyond the reach of our calculations. Also, since $E_z/E$ is very 
small in the later stages of the decay, $l_z$ relies in fact on low 
values of $u_z$ and only reflects a weak component of the flow, from the 
Energy point of view. 
Despite its relative weakness, Ekman pumping is responsible for the larger values of $l_\perp$ 
at $\ha=224$ and more noticeably at $\ha=112$ (figure \ref{fig:lz_f}). Ekman 
pumping indeed transports momentum radially outward of large structure thus 
increasing their effective size in the horizontal plane 
\citep{sommeria88}.\\ 
\begin{figure} \centering
    \input{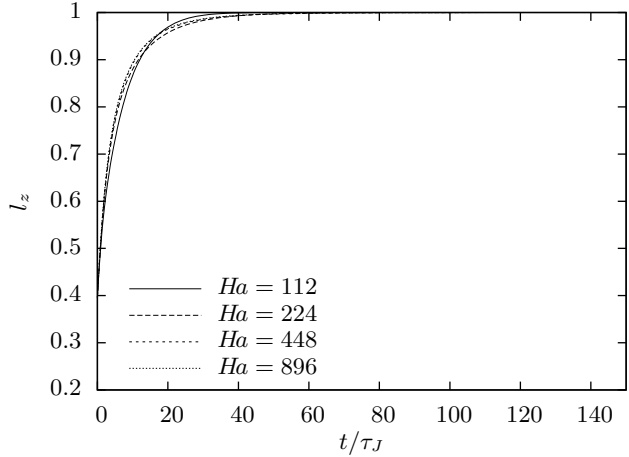}
    \caption{Evolution of integral lengthscale $l_z$ for periodic cases.} \label{fig:lz_f_p}
\end{figure}
%
%\begin{figure}
%\centering
%\includegraphics[angle=-90,width=0.75\textwidth]{uz_prof.ps}
%\caption{Evolution of $u_z$ along vertical line  at $x=y=0$}%coming through the midpoint of the domain.}
%\label{fig:uz_prof}
%\end{figure}
%
\begin{figure} \centering
    \input{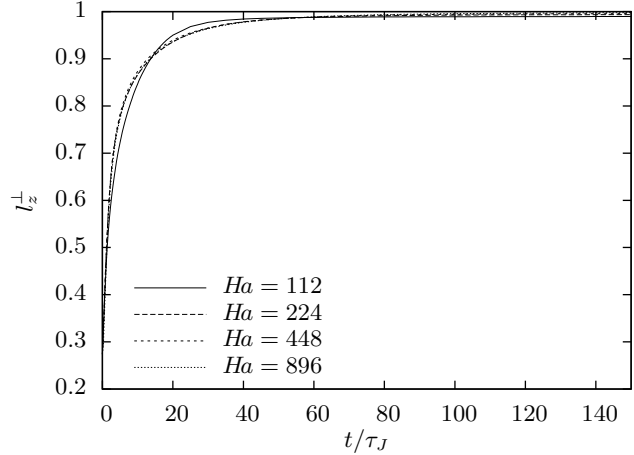}
    \caption{Evolution of integral lengthscale $l_z^\perp$ for simulations with
walls.} \label{fig:lperp_f}
\end{figure}
From this phenomenology, $l_z$ appears dominated by the dynamics of secondary flows but does not reflect accurately the dimensionality of turbulence. We argue that this calls for 
%The influence of this two phenomena  (the \textit{barrel} effect and Ekman
%pumping) on $l_z$ show that there is a need of
 a more suitable quantity to
characterise the growth of vertical scales along the $z$ direction.
%, both for calculations with and without walls. 
We propose that such a quantity may be defined as:
\begin{equation} 
l_z^\perp = \frac{\int\int{\!\mathbf{u}_\perp(x,y,z)
    \mathbf{u}_\perp(x,y,z+z')\,\mathrm{d}V}
    dz'}{\int\!u_\perp^2(x,y,z)\,\mathrm{d}V} \,. 
\label{eq:lzuperp}
\end{equation}
%
%where $\mathbf{u}_\perp$ is the component of velocity perpendicular to the
%magnetic field. 
The temporal evolution of $l_z^\perp$ is shown on fig.
\ref{fig:lperp_f} for decaying turbulence between walls. In contrast to $l_z$ 
it asymptotically converges to 1 for all values of $\ha$ and curves remain very 
close to each other in all stages of the evolution. During the initial stage, 
$l_z^\perp$ grows over a time scale of $\tau_{2D}(l)$, which does reflect the 
two-dimensionalisation of the large scales.  $l_z^\perp$ reaches a value of 0.9 
at $\sim 15 \tJ$ for all values of $\ha$, and its subsequent evolution in the 
quasi-two-dimensional phase is only slow. This indicates $l_z^\perp$ is only 
weakly affected by the two-dimensionalisation of smaller scales, unlike the 
total kinetic energy and the dissipation.
%
%\begin{figure} \centering
%    \input{bigbox}
%    \caption{Evolution of energy and dissipation rates in big box simulations} \label{fig:bigbox_en}
%\end{figure}
%
%\begin{figure} \centering
%    \input{bigboxlen}
%    \caption{Evolution of $l_z$ in big box simulations} \label{fig:bigbox_len}
%\end{figure}
%

\section{Robustness analysis \label{sec:bigbox}}
In order to focus on the long term evolution of the decay of turbulence and 
still keep computational costs reasonable, we have conducted our study with 
a domain of limited size and used only one set of isotropic initial conditions. 
Ongoing experiments on decaying turbulence conducted on the FLOWCUBE setup (\cite{pk14_jfm}) show that reliable quantitative laws require ensemble averaging 
on a large number of initial conditions, which cannot be done numerically.
Nevertheless, we shall now estimate the impact of these choices on the 
mechanisms found, by examining the result of two simulations in a channel four times bigger (dimensions $2L\times2L\times L$), with different random initial 
conditions (albeit with the same statistical properties as in the cases where 
energy has not been boosted), and slightly smaller initial Reynolds number (see table \ref{tab:simul}).\\
\begin{figure} \centering
    \input{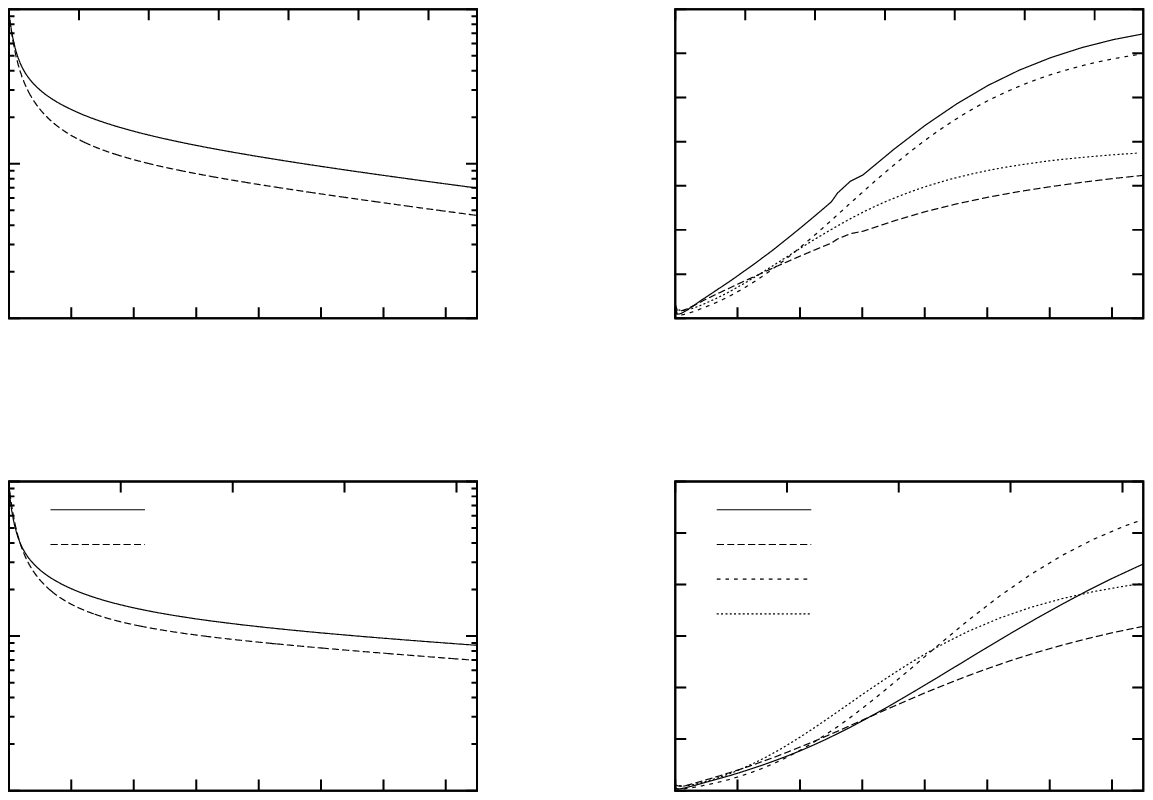}
    \caption{Evolution of energy (left) and dissipation rates (right) in smaller and larger domains (marked "big" in the legend), with Hartmann walls.} \label{fig:bigbox_en}
\end{figure}
\begin{figure} \centering
    \input{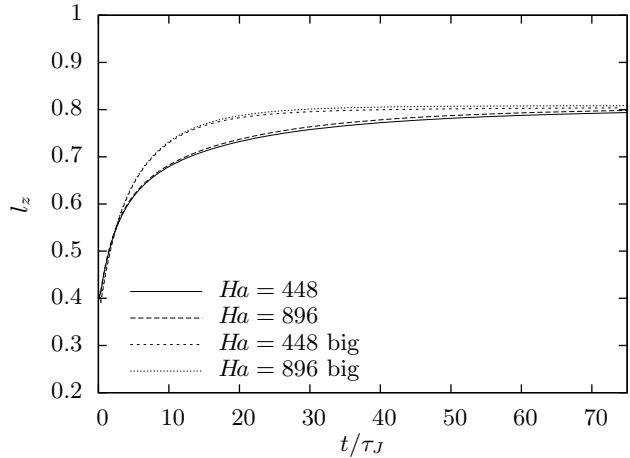}
    \caption{Evolution of $l_z$ in smaller and larger domains (marked "big" in the legend) with Hartmann walls} \label{fig:bigbox_len}
\end{figure}
%
%The two simulation on a box of larger size confirm the dynamics observed on the cubic domain. 
Comparison between energy decay and dissipation ratios in small and large 
domains is shown on figure \ref{fig:bigbox_en}. The decay follows a similar 
profile in both cases. In the three-dimensional phase, 
fits to \cite{okamoto10_jfm}'s decay laws yields exponents that are 
consistently higher in the case of a larger domain. Nevertheless their 
variations with $Ha$ and with the fitting interval are consistent between cases (see table \ref{tab:coefc}).
We also noticed that actual values of the exponent are sensitive to the 
lower bound of the fitting interval, but that again, variations with $Ha$ 
and with the fitting interval are consistent between cases, for the same lower 
bound. This sensitivity can most likely be attributed to the lack of 
an ensemble average which, as in experiments, would be required for a 
precise estimate of exponents in \cite{okamoto10_jfm}'s decay laws. 

Nevertheless, the interval of validity of the these laws, can still be 
estimated by varying the upper bound of the fitting interval, and this 
yields consistent results between cases and choices of fitting interval. 
Most importantly, the main properties 
of the decay outlined throughout the paper appear consistent between 
the two sets of simulations: the duration of the three-dimensional phase is of 
the order of $\tau_{2D}(l_0)$ in both cases, and the asymptotic behaviours are 
identical as decays of energy are parallel to each other. Similarly, 
Ratios of dissipations (dissipation in the Hartmann layer to dissipation in the bulk and viscous to Joule dissipation), depict the same phenomenology as the evolution of energy: timescales are identical for both domains and evolution curves are parallel. The integral lengthscale (figure \ref{fig:bigbox_len}) too evolves initially in a similar fashion in both cases and converges to the same asymptotic value. In conclusion to this short analysis, although it is difficult to 
precisely verify the numerical value of the exponents predicted in 
\cite{okamoto10_jfm}'s laws, the scenario for the decay outlined through the 
analysis cubic domain seems robust to changes in numerical parameters.

%During an intermediate time $\tau_{2D}(l_0)\lesssim t\lesssim 50\tJ$, however, 
%the integral lengthscale is smaller in the simulation with the smaller domain. 
%Despite this difference, timescales and asymptotic behaviour are both 
%domain-independent.
%
%%A possible explanation for this difference may be that energy transfer to 
%%larger scales than in the cubic box may take place that would reduce the 
%%amount of energy fed to smaller scales. Since larger scales become 
%%two-dimensional over a shorter time scale than smaller scalers, this could 
%%explain that the asymptotic value of $l_z$ is reached earlier in the larger 
%%domain.  

\section{Conclusion \label{sec:conclusion}}
Using a new type of spectral methods based on the least dissipative eigenmodes 
of the dissipation operator, we were able to perform direct numerical 
simulations of freely decaying turbulence in a strong magnetic field, between 
two Hartmann walls. The decay exhibits three- and a two-dimensional phases 
with an overlap: the former is dominated 
by the two-dimensionalisation process, where diffusion by the Lorentz force 
stretches vortices until they reach the Hartmann walls. This process is highly 
dissipative and leads to a rapid variation of energy and of the integral 
lengthscale along $\mathbf B$. Larger scales are  two-dimensionalised more 
quickly than smaller ones. Once the large scales of turbulence are close to 
two-dimensional (after $\tau_{2D}(l_0)$) the flow starts exhibiting a 
two-dimensional dynamics, where dissipation mostly takes place in the Hartmann 
boundary layers, with a slower characteristic timescale $\tH=\ha \tJ$. However
since it can take up to $\tH$ for small scales to adopt a two-dimensional dynamics, there is no clear separation between these two phases and both two and three-dimensional dissipation mechanisms co-exist long after $\tau_{2D}(l_0)$. 
We were able to single out several important features of this phenomenology:\\
First, the presence of the walls turned out to impede the growth of $l_z$
right from the earliest stages of the decay, whereas the decay of energy 
remained roughly in line with \cite{okamoto10_jfm}'s law of  $E\sim t^{-1/2}$ for unbounded turbulence in the limit of high $\ha$, during around one Joule time.\\
Second, energy associated to the velocity component across the channel is very strongly suppressed: $E_z/E$ tends to 0 much faster than for unbounded turbulence. This result is consistent with \cite{kolesnikov1974_fd}'s experiments and 
confirms \cite{burattini10_jfm}'s hypothesis that the presence of walls is 
 responsible for the suppression of the third component. Further evidence of 
this suppression is visible in the long term behaviour of the skewness which tends to 0 in the case with walls. With periodic boundary conditions, by contrast, 
the Skewness apparently tends to a  constant value. However, this only seems 
true at high $Ha$. Since, two-dimensionalisation occurs over a timescale of 
the order of $\tH$ which is much longer than our calculations at high $Ha$, 
 and than calculations in previous studies, it is unclear whether the Skewness 
indeed remains constant past the two-dimensional phase with periodic 
boundaries.\\
Long into the "two-dimensional phase", we found that even at the highest value 
of $\ha$, a form of three-dimensionality subsisted, due to currents recirculating between the Hartmann layers and the bulk. This effect is characterised by the 
barrel shape visible on the larger structures, as predicted by \cite{psm00_jfm}.
Though less pronounced at higher values of $\ha$, our simulations show no 
evidence of it vanishing at larger times.\\
Thirdly, in quasi-two dimensional flows dominated by dissipation in the 
Hartmann boundary layers, the total kinetic energy would be expected to decay 
exponentially 
with a timescale of $\tH$.  However, a true exponential decay of this sort was 
never observed, even for $t\gtrsim\tH$. Remarkably this discrepancy to a pure 
exponential decay did not result from the residual three-dimensionality due to 
the barrel effect, but mostly from viscous friction in the horizontal plane.\\
Finally, at more moderate values of $\ha$ ($\ha=112$), secondary flows in large 
structures significantly affect the decay by increasing the integral 
lengthscale in the directions along the channel (perpendicular to $\mathbf B$). Since, however this effect is driven by two-dimensional inertia, it is expected 
to vanish at larger times, but was still present after $2.5\tH$. 
When present, it is shown to dominate the behaviour of the integral lengthscale in the direction of the magnetic field. This prompted us to put forward an 
alternative definition for this integral lengthscale that gives a better 
measure of the flow dimensionality.\\

The authors are grateful to the Leverhulme Trust, who supported this work through Research Project Grant ref. F00/732J and to Professor Mahendra K. Verma, who made his Fourrier-base spectral code available to them as a basis for the implementation of their novel spectral method.

\appendix
%\textcolor{blue}{
%\input{turbu_fluc.tex}
%\input{correlation_noise.tex}
%}
\bibliographystyle{jfm}
\bibliography{fullbiblio.bib}

\end{document}